\documentclass{aa}

\usepackage{txfonts}
\usepackage{supertabular}
\usepackage{rotating}
\usepackage{dsfont}
\usepackage{graphicx}
\usepackage{epstopdf}
\usepackage{color}
\usepackage{tabularx}
\usepackage{calc}
\usepackage{array}
\usepackage{natbib}
\bibpunct{(}{)}{;}{a}{}{,} 

\begin{document}

\title{Identification of galaxy clusters in cosmic microwave background maps using the Sunyaev-Zel'dovich effect}

\author{C. P. Novaes \and C. A. Wuensche}

\institute{Divis\~ao de Astrof\'isica, Instituto Nacional de Pesquisas Espaciais - INPE, S\~ao Jos\'e dos Campos, SP, Brazil}

\offprints{C. P. Novaes \email{camila@das.inpe.br}}

\date{Recieved 18 November 2011 /Accepted 25 July 2012}

\abstract  {The \textit{Planck} satellite was launched in 2009 by the European Space Agency to study the properties of the cosmic microwave background (CMB). An expected result of the \textit{Planck} data analysis is the distinction of the various contaminants of the CMB signal. Among these contaminants is the Sunyaev-Zel'dovich (SZ) effect, which is caused by the inverse Compton scattering of CMB photons by high energy electrons in the intracluster medium of galaxy clusters.}
{We modify a public version of the JADE (Joint Approximate Diagonalization of Eigenmatrices) algorithm, to deal with noisy data, and then use this algorithm as a tool to search for SZ clusters in two simulated datasets.} 
{The first dataset is composed of simple ``homemade" simulations and the second of full sky simulations of high angular resolution, available at the LAMBDA (Legacy Archive for Microwave Background Data Analysis) website.
The process of component separation can be summarized in four main steps: (1) pre-processing based on wavelet analysis, which performs an initial cleaning (denoising) of data to minimize the noise level; (2) the separation of the components (emissions) by JADE; (3) the calibration of the recovered SZ map; and (4) the identification of the positions and intensities of the clusters using the SExtractor software.}
{The results show that our JADE-based algorithm is effective in identifying the position and intensity of the SZ clusters, with the purities being higher then 90\% for the extracted ``catalogues". This value changes slightly according to the characteristics of noise and the number of components included in the input maps.}
{The main highlight of our developed work is the effective recovery rate of SZ sources from noisy data, with no \emph{a priori} assumptions. This powerful algorithm can be easily implemented and become an interesting complementary option to the ``matched filter'' algorithm (hereafter MF) widely used in SZ data analysis.}


\titlerunning{Identification of galaxy clusters in CMB maps using the Sunyaev-Zel'dovich effect}
\authorrunning{Novaes and Wuensche}
\maketitle


\section{Introduction}

During the passage of the cosmic microwave background (CMB) radiation through clusters of galaxies, about 1\% of the photons are Compton scattered by energetic electrons in the intracluster medium. This process causes a very distinctive signature in the CMB spectrum, that was first described by \citet{sz/1969}.

The Sunyaev Zel'dovich (SZ) effect is a secondary CMB anisotropy, meaning that it was produced after the decoupling era. Its angular size is of the order of arc minutes and an average intensity of a few hundred  $\mu K$, which is difficult to separate from the primary CMB signal and therefore difficult to detect. However, some currently operating ground-based experiments, such as the South Pole Telescope (SPT) and the Atacama Cosmology Telescope (ACT), have sufficiently high sensitivities to measure the SZ effect with high signal-to-noise ratio data and enough angular resolution to obtain a very accurate SZ profile from the observed clusters \citep{2011act}.

Together with current optical and X-ray surveys, SZ measurements are expected to produce cluster images with the highest possible sensitivities across significant fractions of the sky \citep{2011carlstrom,2011marriage,2011/planck_results_cluster}. The multiwavelength data will be used to shed light on the cluster physics, to improve our knowledge of scaling relations, and to produce catalogues to be used in cosmological studies. Measurements of the SZ effect offers a unique and powerful tool to test cosmological models and put strong constraints on the parameters describing the universe \citep[e.g.,][]{2005/voit, 2011/allen}. In addition to the SZ effect, the hot intracluster gas is also characterized by its strong bremsstrahlung emission in the X-ray band. Put together, both effects can be used to estimate the distance of clusters and the Hubble constant. In addition, the SZ effect can also be used to estimate the $\Omega_B / \Omega_M$ ratio and the peculiar velocity of clusters. One can also use large SZ surveys to constrain  the dark energy equation of state  \citep[see, e.g.,][]{birkinshaw, carlstromb, carlstrom}.

A full sky survey is being conducted by the \textit{Planck} satellite, launched in 2009 and the first mission of the European Space Agency (ESA) dedicated to CMB studies. In January 2011, the \textit{Planck} team released the first version of its full-sky SZ cluster catalogue \citep{2011/planck_results_cluster}. These results are already being used to study the CMB contamination on angular scales smaller than a few arcminutes ($\ell \gtrsim 1000$), where the SZ effect, together with radio and sub-mm point sources, dominate over the primary CMB contribution \citep[e.g.,][]{2010/taburet}.

A number of algorithms have been used to extract SZ signal from CMB maps, but most use \emph{a priori} assumptions about the SZ signal contained in the input maps and identify the ``unknown'' clusters based upon spectral identification and information about shape, intensity, etc \citep[see, e.g.,][]{Boomerang2003, ACBAR2005, 2008bobin, leach, vanderlinde2010, ILC2011}.

The main purpose of this work is to present a method to identify SZ clusters in CMB maps, using a minimal set of \emph{a priori} conditions. To do this, we developed a ``\emph{blind} search'' method, based only on the spectral contributions of input signals, that has performed very well in simulated sky maps with include many of the public \textit{Planck} satellite characteristics, such as the asymmetric sky coverage, detector noise level, frequency coverage, etc.

The outline of this paper is as follows: in Section \ref{sz}, we briefly describe the SZ effect theory and the pressure profile described by M. Arnaud and collaborators. Section \ref{simulation} contains the details of the two datasets used in this work, one composed of ``homemade" simulations and another produced by \citet{2010/sehgal}. The methodology used to identify SZ clusters is discussed in Section \ref{simulation}. Section \ref{results} summarizes our results and our concluding remarks are presented in Section \ref{closing}.


\section{The Sunyaev Zel'dovich effect}\label{sz}

The SZ effect produces a small distortion in the CMB spectrum, with a temperature variation $\Delta T_ {SZ}$ given by

\begin{equation} \label{dist_tot}
\frac{\Delta T_{SZ}}{T_{CMB}} = f(x) y - \tau_e \bigg(\frac{v_{pec}}{c} \bigg).
\end{equation}

\noindent The first term in Eq. \ref{dist_tot} corresponds to the distortion caused by the thermal distribution of electrons located in the intracluster medium that scatter the CMB photons. The Comptonization parameter $y$  is given by $y = \int \bigg(\frac{k_B T_e}{m_e c^2}\bigg) \sigma_T n_e dl$, where $\sigma_T$ is the Thomson cross-section, $n_e$ the electron density, $dl$ the line element along the line of sight, and $f(x)$ the frequency dependence given by

\begin{equation} \label{dep_freq}
f(x) = \bigg(x \frac{e^x + 1}{e^x - 1} - 4 \bigg) (1 +
\delta_{SZ}(x,T_e)),
\end{equation}

\noindent where $x = h \nu / k_B T_{CMB}$ and $\delta_{SZ}(x,T_e)$ is the relativistic correction.

The second term in Eq. \ref{dist_tot}, the so-called kinetic SZ effect, refers to the spectral distortion caused by the movement of the cluster relative to the CMB radiation. It is caused by the cluster speed, which creates a Doppler distortion of the scattered photons, with $\tau_e$ being the optical depth, $v_{pec}$ the speed of the cluster towards the line of sight, and $c$ the speed of light. This work considers both the thermal and kinetic contributions in the synthetic maps produced by \citet{2010/sehgal} and only the thermal contribution in our own simulations, since the thermal effect is usually at least one order of magnitude larger than the kinetic one.

Equation \ref{dist_tot} can be rewritten to take into account the variation in the Comptonization parameter $y$ as a function of the radial coordinate of the projected cluster, following the discussion in \citet{wmapsz}

\begin{equation}
\frac{\Delta T_{SZ}}{T_{CMB}}(\theta) = f(x) \frac{\sigma_T}{m_e c^2} \int^{l_{out}}_{-l_{out}} P_e\bigg(\sqrt{l^2 + \theta^2D_A^2}\bigg) dl,
\end{equation}

\noindent where $\theta$ is the angular distance from the cluster centre, $D_A$ the angular diameter distance, $l$ the radial coordinate from the centred of the cluster along the line of sight, $\sigma_T$ the Thomson cross-section, $m_e$ the electron mass, $c$ the speed of light, and the electron pressure profile is given by $P_e = n_e k_B T_e$. For a given pressure profile $ P_e (r) $, the SZ temperature variation $\Delta T_{SZ}$ can be written as

\begin{equation}
\Delta T_{SZ}(\theta) = f(x) T_{CMB} \frac{\sigma_T}{m_e c^2} P^{2d}_e(\theta),
\end{equation}

\noindent where $P^{2d}_e(\theta)$ is the electron pressure profile projected in the sky given by

\begin{equation}
P^{2d}_e(\theta) = \int^{\sqrt{r_{out}^2 - \theta^2 D^2_A}}_{-\sqrt{r_{out}^2 - \theta^2 D^2_A}} P_e\bigg(\sqrt{l^2 + \theta^2D_A^2}\bigg) dl.
\end{equation}

\noindent Here, the pressure profile is truncated in $r_{out}$.  \citet{2010/arnaud} define an electron pressure profile $P_e$, based on the generalized Navarro-Frenk-White (NFW, \cite{1997NFW}) model described by \citet{nagai_nfw}. This profile closely describes the electron pressure profile obtained from X-ray data, and is given by

\begin{eqnarray}
P_e(r)&=&1.65 \times 10^{-3} h(z)^{8/3} \bigg[\frac{M_{500}}{3 \times 10^{14} h^{-1}_{70} M_{\odot}}\bigg]^{2/3 + \alpha_p} \nonumber\\ & & \times ~ p(x) h^2_{70}~keV cm^{-3},
\end{eqnarray}

\noindent where $h(z)$ is given by $h(z) = [\Omega_m (1+z)^3 + \Omega_{\Lambda}]^{1/2}$  and is the ratio of the Hubble constant at redshift z to its present value, $H_0$. Moreover, $\alpha_p = 0.12$ and $x=r/R_{500}$, where $R_{500}$ is the radius within which the mean overdensity is 500 times the critical density of the universe at redshift $z$ ($\rho_c(z) = 2.775 \times 10^{11} E^2(z) h^2 M_{\odot} Mpc$), $M_{500}$ is the mass within the radius $R_{500}$, given by

\begin{equation}
M_{500} \equiv \frac{4 \pi}{3}[500 \rho_c(z)]R_{500}^3,
\end{equation}

\noindent $p(x)$ corresponds to the generalized NFW model

\begin{equation}
p(x) = \frac{P_0}{(c_{500} x)^{\gamma}[1 + (c_{500} x)^{\alpha}]^{(\beta - \gamma)/\alpha}},
\end{equation}

\noindent and the best-fit found by \citet{2010/arnaud} is given by

\begin{eqnarray}
[P_0,c_{500},\gamma,\alpha,\beta] &=& [8.403 h^{-3/2}_{70},1.177,0.3081,\nonumber\\
                                  & & 1.0510,5.4905].
\end{eqnarray}


\section{Description of simulated data}\label{simulation}

To perform a thorough testing of our method, we used two different sets of simulations. The first group is a ``homemade" dataset composed of five components (CMB, SZ effect, synchrotron, free-free, and dust emission) at the frequencies of 100, 143, 217, 353, and 545 GHz (five \textit{Planck} HFI frequencies).  The second group is a more realistic set of sky maps, including, in addition to the aforementioned components, point sources. These simulated data were developed to test the data reduction pipeline for the Atacama Cosmology Telescope (ACT). These maps are available at the LAMBDA\footnote{http://lambda.gsfc.nasa.gov/toolbox/tb\_cmbsim\_ov.cfm} (Legacy Archive for Microwave Background Data Analysis) website. The details of our simulations are presented below, along with a summary of the second set of high-resolution sky simulations.


\subsection{``Homemade" simulations}\label{homemade}

Our goal was to generate simple synthetic maps to reproduce, in the simplest way for an outsider to the \textit{Planck} Collaboration, the observations made by the \textit{Planck} satellite.

All maps were simulated using the HEALPix (Hierarchical Equal Area iso-Latitude Pixelization) pixelization grid \citep{gorski}. The maps produced by \textit{Planck} will have $N_{side}=2048$, which means that each map will consist of $N_{pix} \approx 5 \times 10^7$ pixels of size 1.7 arcmin.

However, as the angular resolutions of the \textit{Planck} instruments for the simulated frequencies at which we simulated the maps are between 10' and 4', it was unnecessary to simulate these maps with pixels of 1.7' in diameter, since this would be about of three times higher spatial resolution than \textit{Planck}'s.

We therefore created maps with $N_{side}=1024$ ($N_{pix} \approx 1.2 \times 10^7$) and average angular diameters of 3.43', which have lower resolutions than the quoted figures for the \textit{Planck} frequencies.
Higher resolutions would enhance the SZ features of the profiles, but since we search for previously unidentified clusters, instead of studying the characteristics of the cluster profile, this does not add significantly to the search process. Moreover, it increases the processing time by a factor of $\sim \Delta N_{pix}^2$, with $N$ being the number of pixels in the map. This set of maps were constructed at the frequencies of 100, 143, 217, 353, and 545 GHz.  A description of the components used in the simulated maps are presented in the following subsections.

\subsubsection{Cosmic microwave background anisotropies}

\begin{figure}[h]
\resizebox{\hsize}{!}{\includegraphics{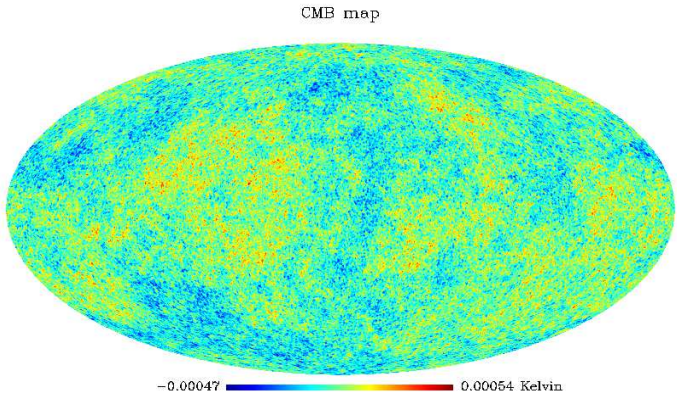}}
\caption{Cosmic microwave background anisotropy map in Mollweide projection, Galactic coordinates, and Kelvin temperatures.} \label{cmb_simul}
\end{figure}

We performed our simulations of the temperature fluctuations in the CMB based on the $C_l$ coefficients created using the online interface of CMBFAST code \citep{seljak}. We considered the $\Lambda CDM$ standard model with $\Omega_M \sim 0,27$, $\Omega_{\Lambda} \sim 0,73$, $\Omega_b h^2 \sim 0,024$, and $h = 0,72$. From this spectrum, the field of CMB primary anisotropies of the whole sky was generated using the SYNFAST routine of the HEALPix package. Figure \ref{cmb_simul} shows the synthetic CMB map, in thermodynamic temperature.

\subsubsection{Galactic emission}

The Galactic contribution to the synthetic maps was added using the WMAP 7-year (hereafter WMAP-7) ``derived foreground products" maps  \citep{wmap7, wmap/foreg} available at the LAMBDA website. However, the WMAP measurement frequencies are different from those used in this work and we had to scale the emission maps to the \textit{Planck} frequencies, assuming that they follow a power law with indexes estimated by the WMAP team.  The intensity $I_e$ of each Galactic emission $e$, with spectral index $\beta_e$, depends on the frequency $\nu$ according to \citep{wmap/first}

\begin{equation}
  I_e(\nu) \propto \nu^{\beta_e}.
\end{equation}

\noindent Since $I_e(\nu_1)$ and $I_e(\nu_2)$ are the intensities of a given emission $e$ at two different emission frequencies ($\nu_1$ and $\nu_2$), you can write the ratio of these intensities as

\begin{equation} \label{calc_mapa}
  \frac{I_e(\nu_1)}{I_e(\nu_2)} = \bigg(\frac{\nu_1}{\nu_2}\bigg)^{\beta_e} \Rightarrow I_e(\nu_1) = I_e(\nu_2) \bigg(\frac{\nu_1}{\nu_2}\bigg)^{\beta_e}.
\end{equation}

\noindent Thus, we used a $I_{\nu_2}$ map of a foreground component at a given frequency $\nu_2$ and the corresponding spectral index, to obtain a synthetically scaled map $I_{\nu_1}$ of emission $e$.

The equation \ref{calc_mapa} was applied, pixel-by-pixel, to maps of synchrotron, dust, and free-free emissions in W band (94 GHz). We did not scale the maps using a pixel-by-pixel fit but instead fixed spectral indices, where $\beta_s = -3,0$, $\beta_d = 2,0$, and $\beta_{ff} = -2,16$ \citep{wmap/foreg} for the three types of emission, respectively. Both the maps of Galactic emission and the spectral index values used are part of the WMAP-7 products and results.

\subsubsection{The SZ effect}

\begin{figure*}[ht!]
\centering
\includegraphics[height=6.5cm]{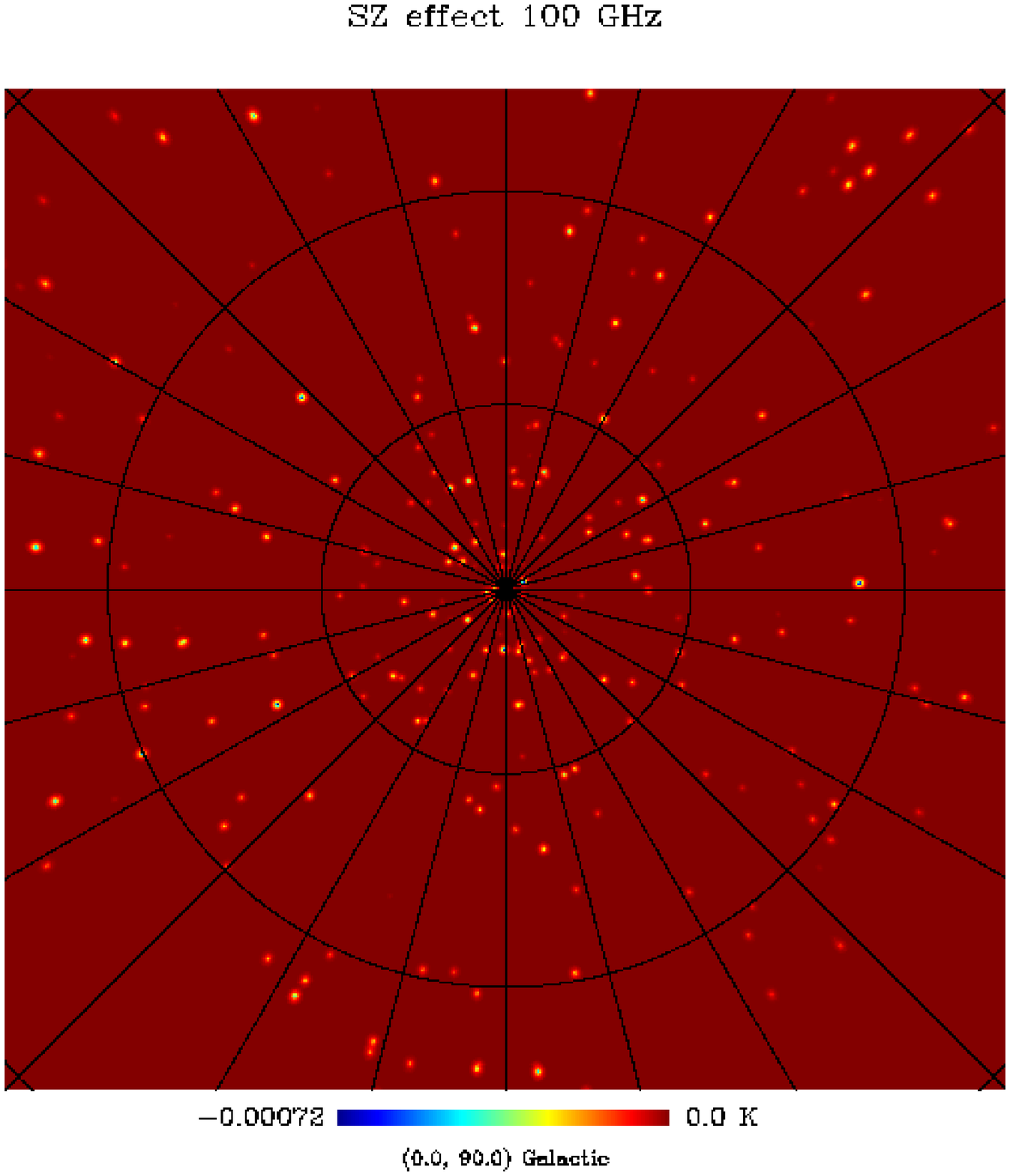}
\includegraphics[height=6.5cm]{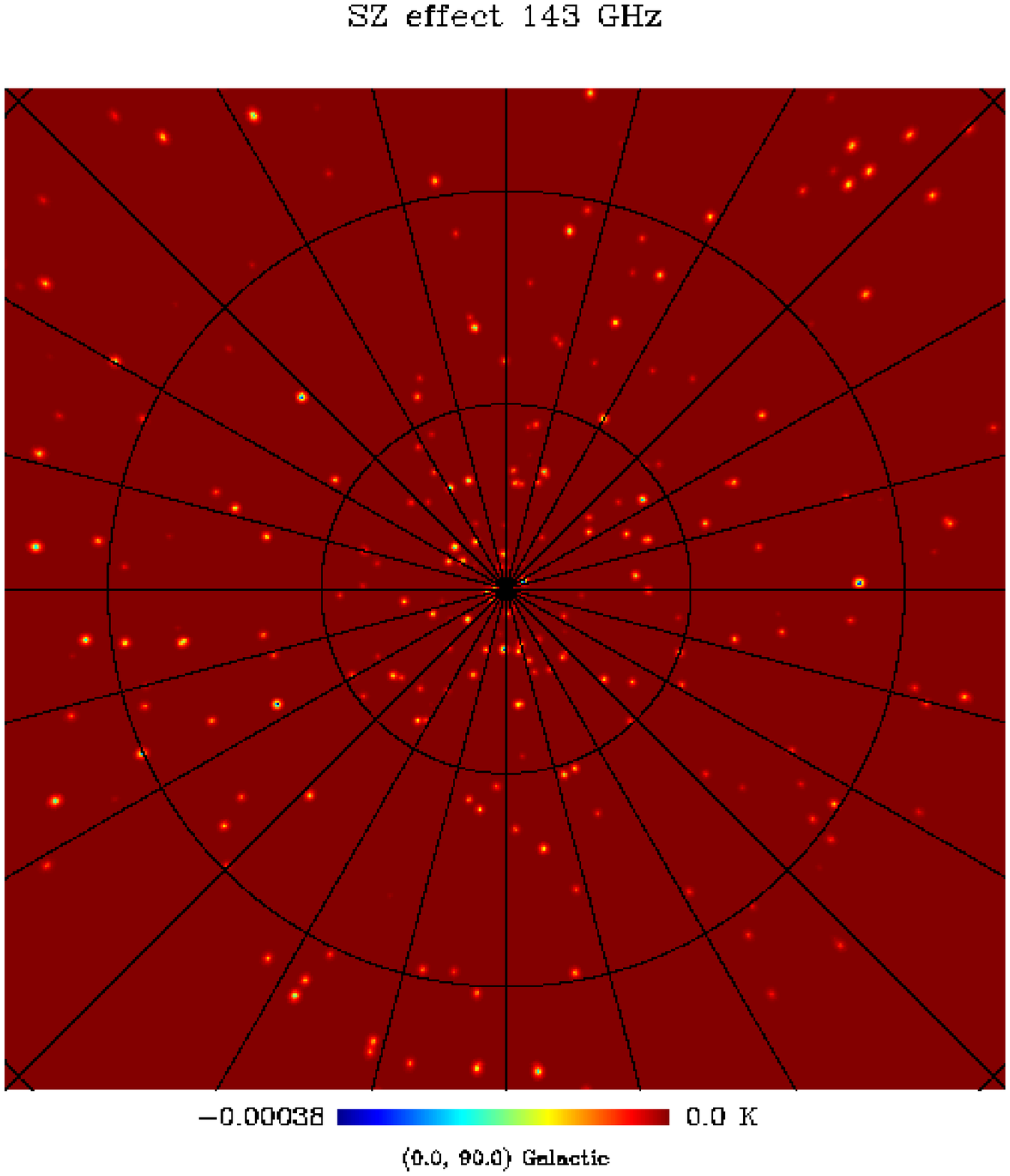}
\includegraphics[height=6.5cm]{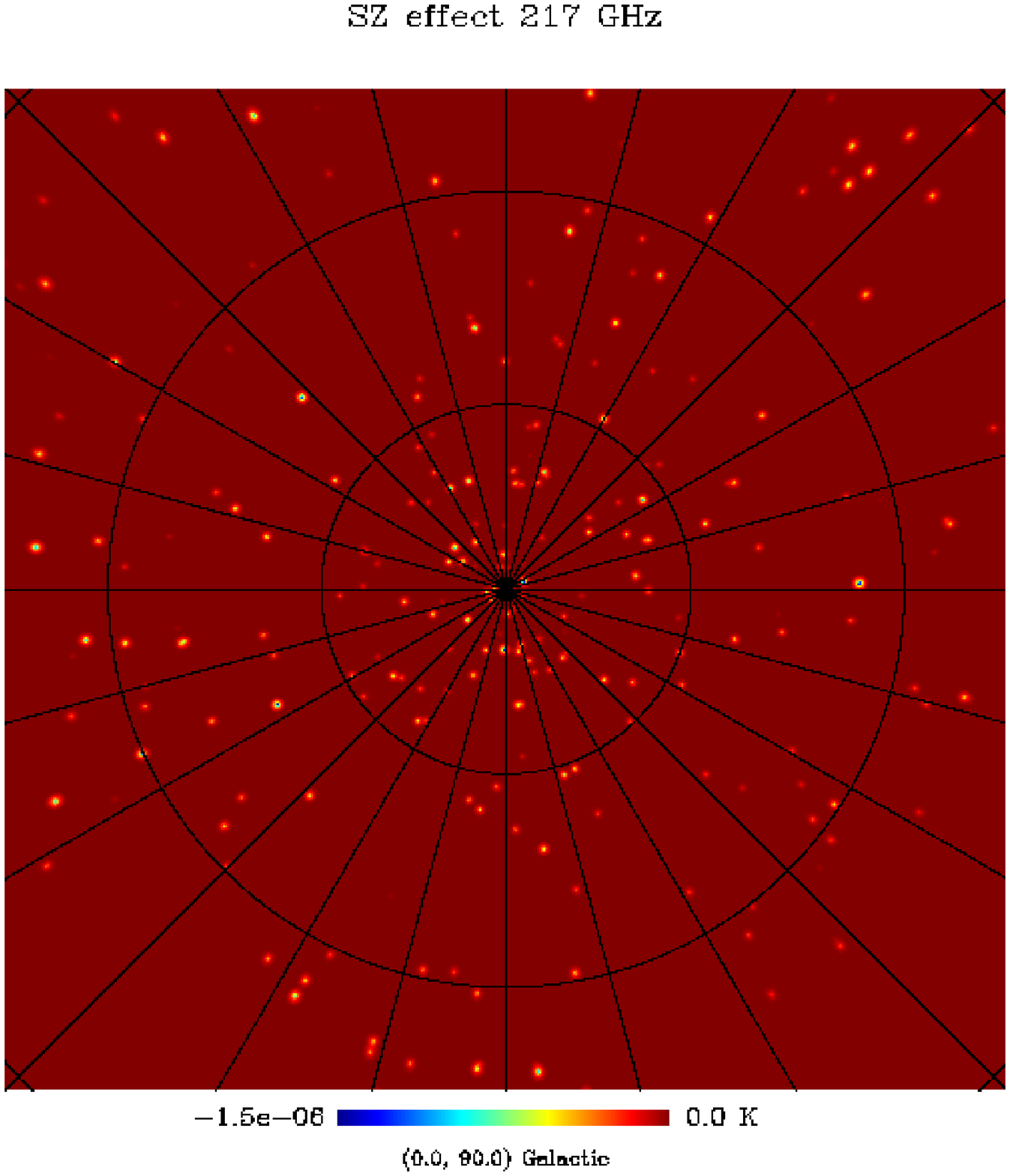}
\includegraphics[height=6.5cm]{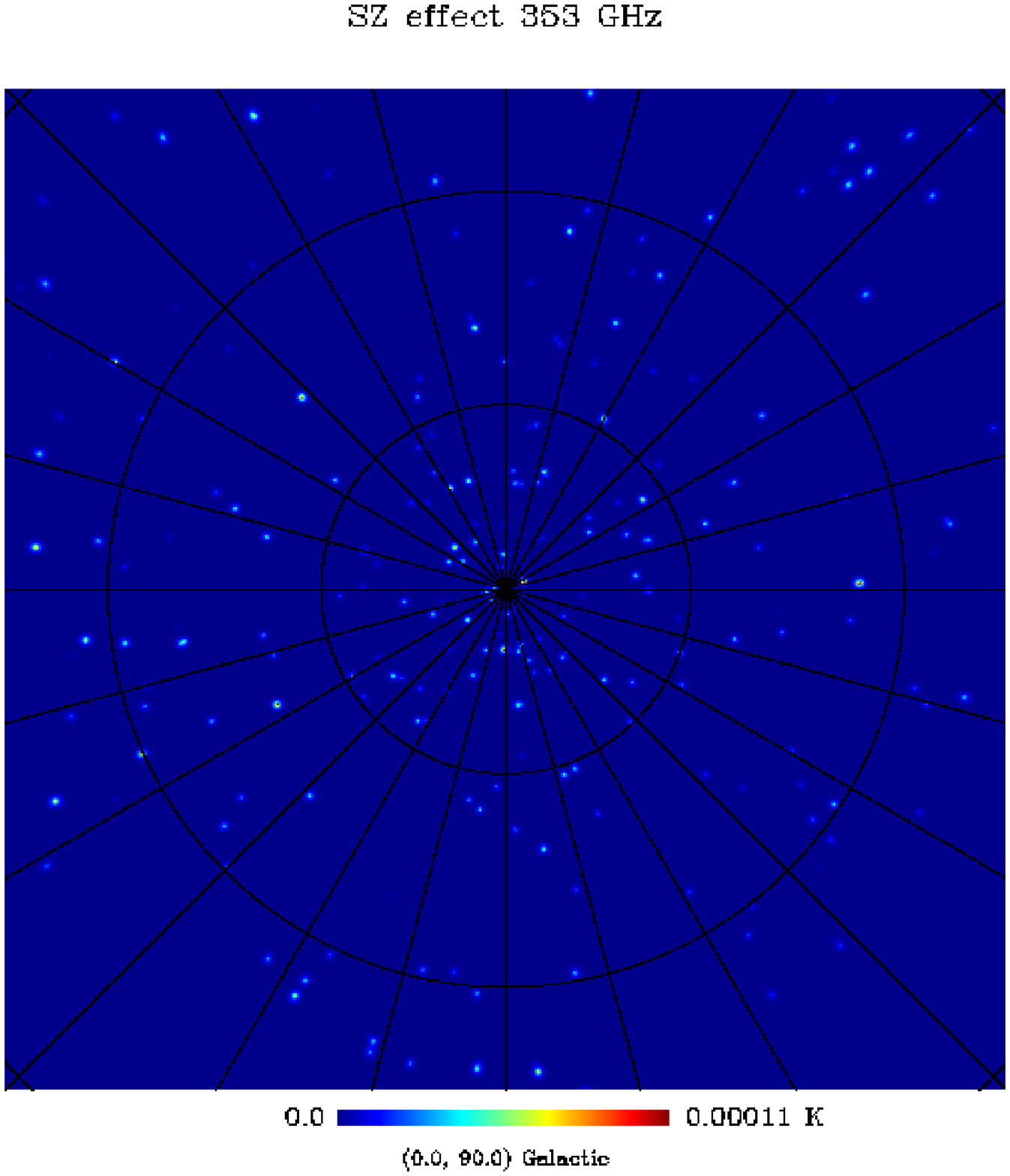}
\includegraphics[height=6.5cm]{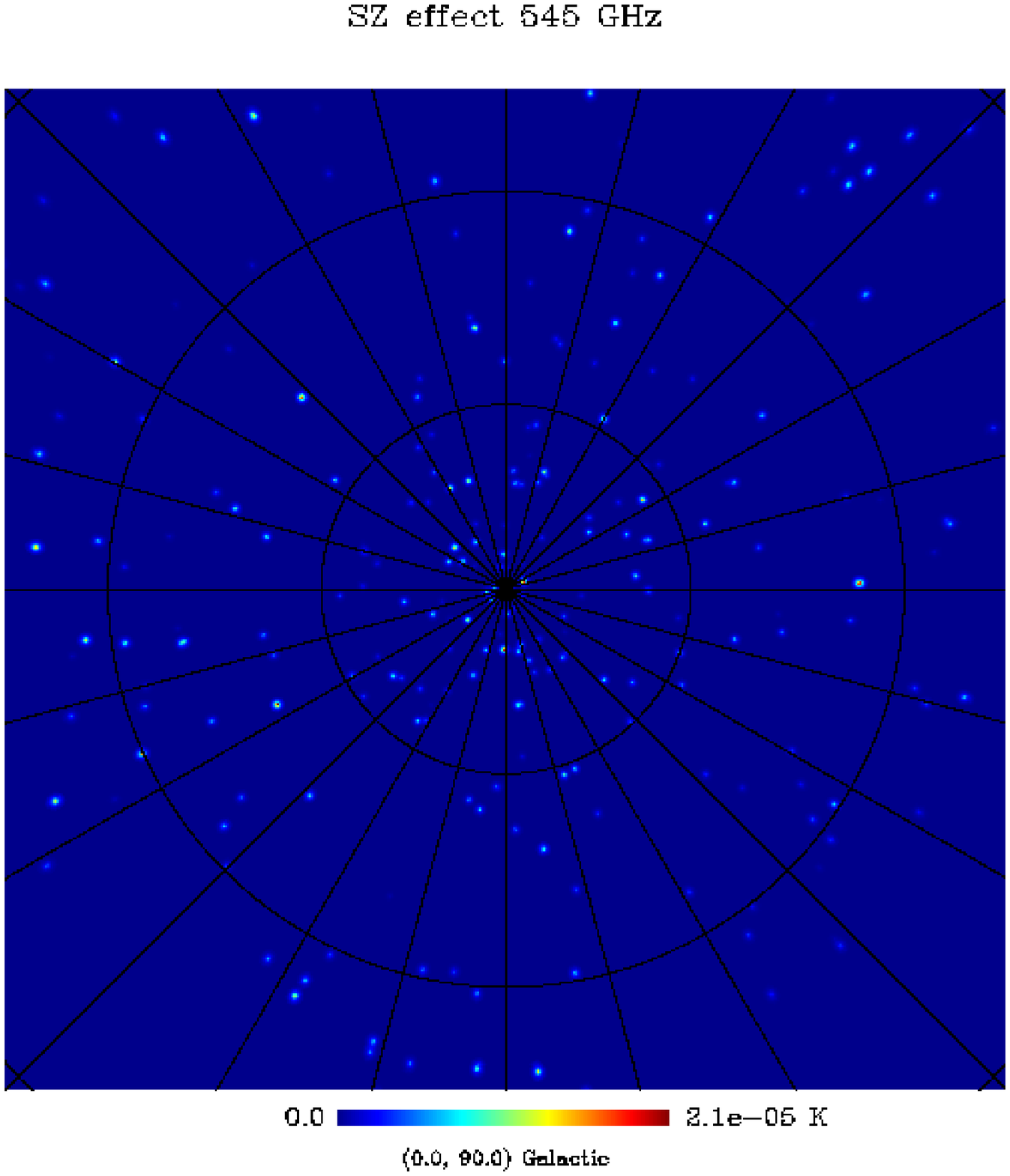}
\caption{Gnomonic projection centered on the north Galactic pole and in Kelvin temperature, of SZ effect in 100, 143, 217, 353 and 545 GHz.} \label{sz_simul}
\end{figure*}

The clusters were produced from the SZ temperature profiles corresponding to the generalized Navarro-Frenk-White model for the pressure profile of the intracluster gas, as described in Section \ref{sz}, using the value $r_{out} = R_{500}$ for the integration. We simulated 1000 synthetic clusters positioned throughout the sky and outside the Galactic region, with random orientations and following a uniform distribution. The temperature profiles were constructed using an adaptation of the routine available at Eiichiro Komatsu's  website\footnote{http://gyudon.as.utexas.edu/$\sim$komatsu/CRL/index.html}, considering mass values $5 \times 10^{13} M_{\odot} < M_{500} < 1 \times 10^{15} M_{\odot}$ and a redshift interval $3 \times 10^{-4} < z < 1.5$. The resulting simulated maps, in the five selected frequencies,  are shown in Fig. \ref{sz_simul}. A section around the north galactic pole was selected to provide a clearer view of the SZ signature at all of the five frequencies.

\subsubsection{Symmetric and asymmetric noise} \label{simul_noise}

The noise was simulated using the white noise sensitivities of each chosen \textit{Planck} channel, which is given in thermodynamic CMB temperature units, estimated for the \textit{Planck} mission (Tab. \ref{bandas}). The simulation was carried out to obtain a map of white Gaussian noise, by assuming both a roughly homogeneous coverage of the sky, and an asymmetric sky coverage mimicking a \textit{Planck} observing scheme.

In the first case (of homogeneous white Gaussian noise, hereafter HWGN), we generated for each frequency, a Gaussian random distribution of zero mean and standard deviation given by the corresponding white noise sensitivity for 15 months of the mission.

In the second case, the \textit{Planck}-like noise due to the asymmetric sky coverage (hereafter NASC) was estimated using the same white noise sensitivities and a scaled version of the observation number ($N_{obs}$) map of WMAP-7, which is also available at the LAMBDA website. Since the most frequently observed regions by both satellites are the ecliptic poles, we constructed a $N_{obs}$ map that we considered an acceptable approximation for \textit{Planck} coverage. The WMAP-7 $N_{obs}$ map was adapted to match \textit{Planck} values and the ``ring" effect around the ecliptic poles, which is not present in the  \textit{Planck} $N_{obs}$ maps, was smoothed out.

Using the above-mentioned $N_{obs}$ map for 15 months and a Gaussian random distribution with zero mean and standard deviation given by the \textit{Planck} white noise sensitivity (in $\mu K s^{1/2}$), we created \textit{Planck}-like noise maps (NASC) for each frequency.

\begin{table}[!h]
\centering
\caption{Characteristics of the \textit{Planck}
satellite instruments (adapted from \citet{2011/planck_collaboration}).} {\footnotesize
\begin{tabular}{c c c c c c}
\hline \hline
Frequency (GHz) & 100 & 143 & 217 & 353 & 545 \\
\emph{FWHM} (arcmin) & 9,37 & 7,04 & 4.68 & 4.43 & 3.80 \\
Sensitivity $^a$ ($\mu K_{CMB} s^{1/2}$) & 22.6 & 14.5 & 20.6 & 77.3 & 1011.3 $^b$ \\
\hline
\multicolumn{6}{p{8cm}}{$^a$ Uncorrelated noise in 1 s for the corresponding array of detectors in each frequency.} \\
\multicolumn{6}{p{8cm}}{$^b$ Obtained from the extrapolation of lower frequencies.}
\end{tabular}} \label{bandas}
\end{table}

\subsubsection{Construction of a simulated \textit{Planck} sky}

Using the components described above, we produced a ``homemade" \textit{Planck} sky. The maps were produced at 100, 143, 217, 353, and 545 GHz from a combination of maps of CMB, SZ effect, synchrotron, dust, and free-free emissions, together with noise, added with equal weights, as described in Equation \ref{combinacao_linear}

\begin{equation} \label{combinacao_linear}
  X^{\nu} = \sum\limits_{i=1}^{N_c} x_i^{\nu},
\end{equation}

\noindent where $x_i^{\nu}$ is the map of the component (emission) $i$ at a given frequency $\nu$ and $X^{\nu}$ is the resulting map of the linear combination of $N_c$ components. Each frequency map was convolved with the corresponding beam, using the full width at half maximum (FWHM) values for the \textit{Planck} channels (Tab. \ref{bandas}), then a realization of the noise was added to each map.

\begin{figure*}[htb]
\centering
\includegraphics[height=5cm]{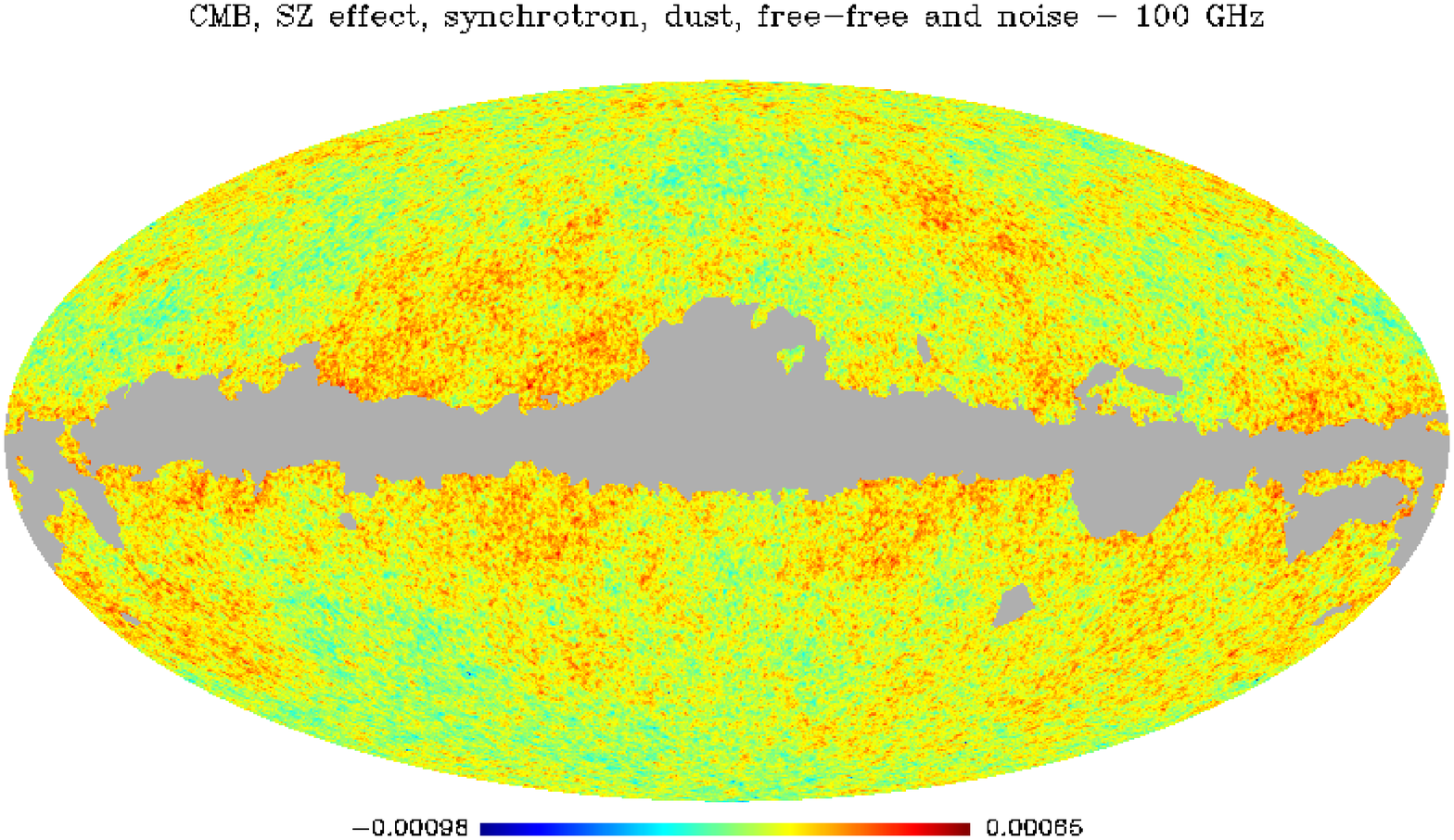}
\includegraphics[height=5cm]{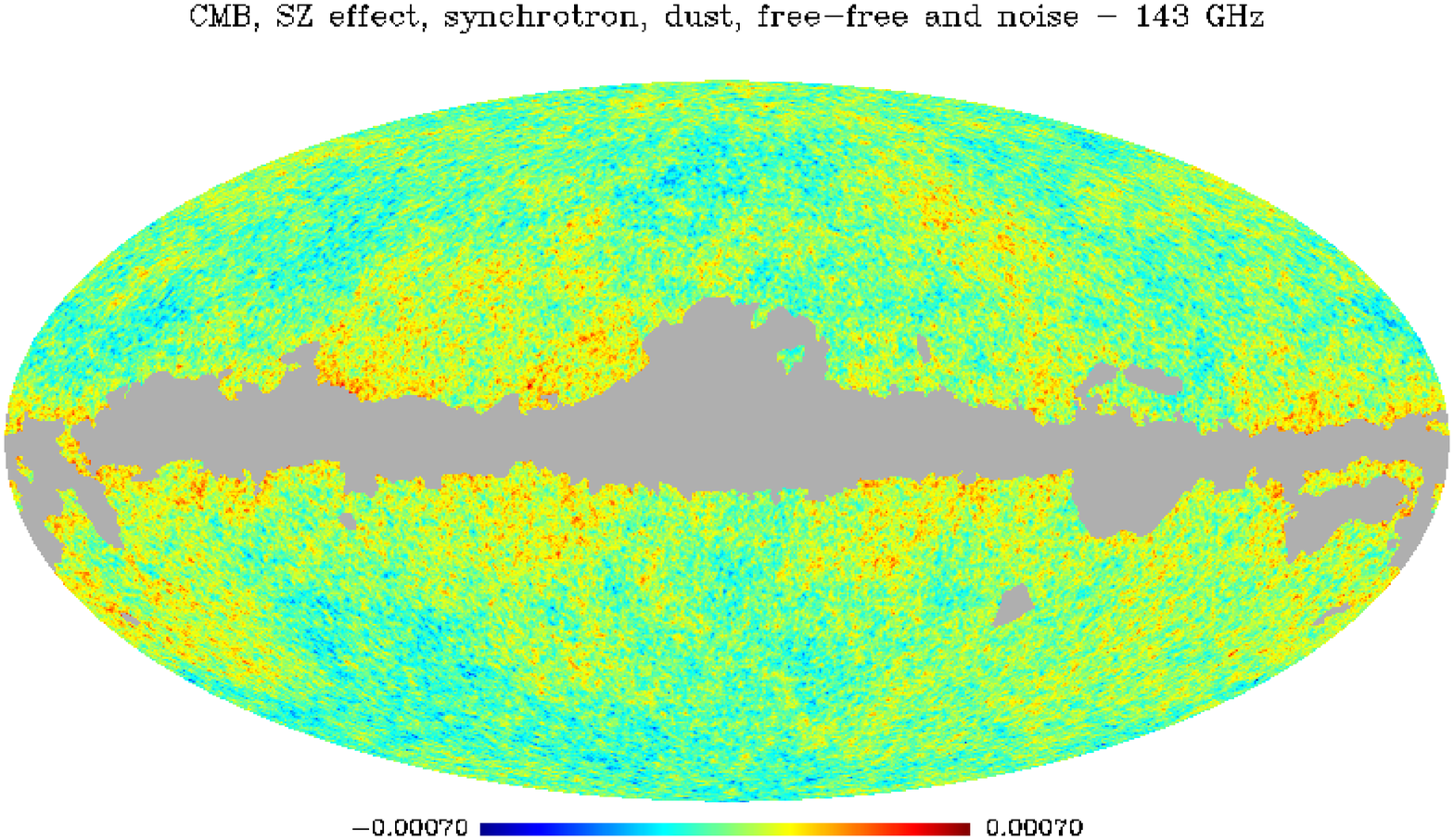}
\includegraphics[height=5cm]{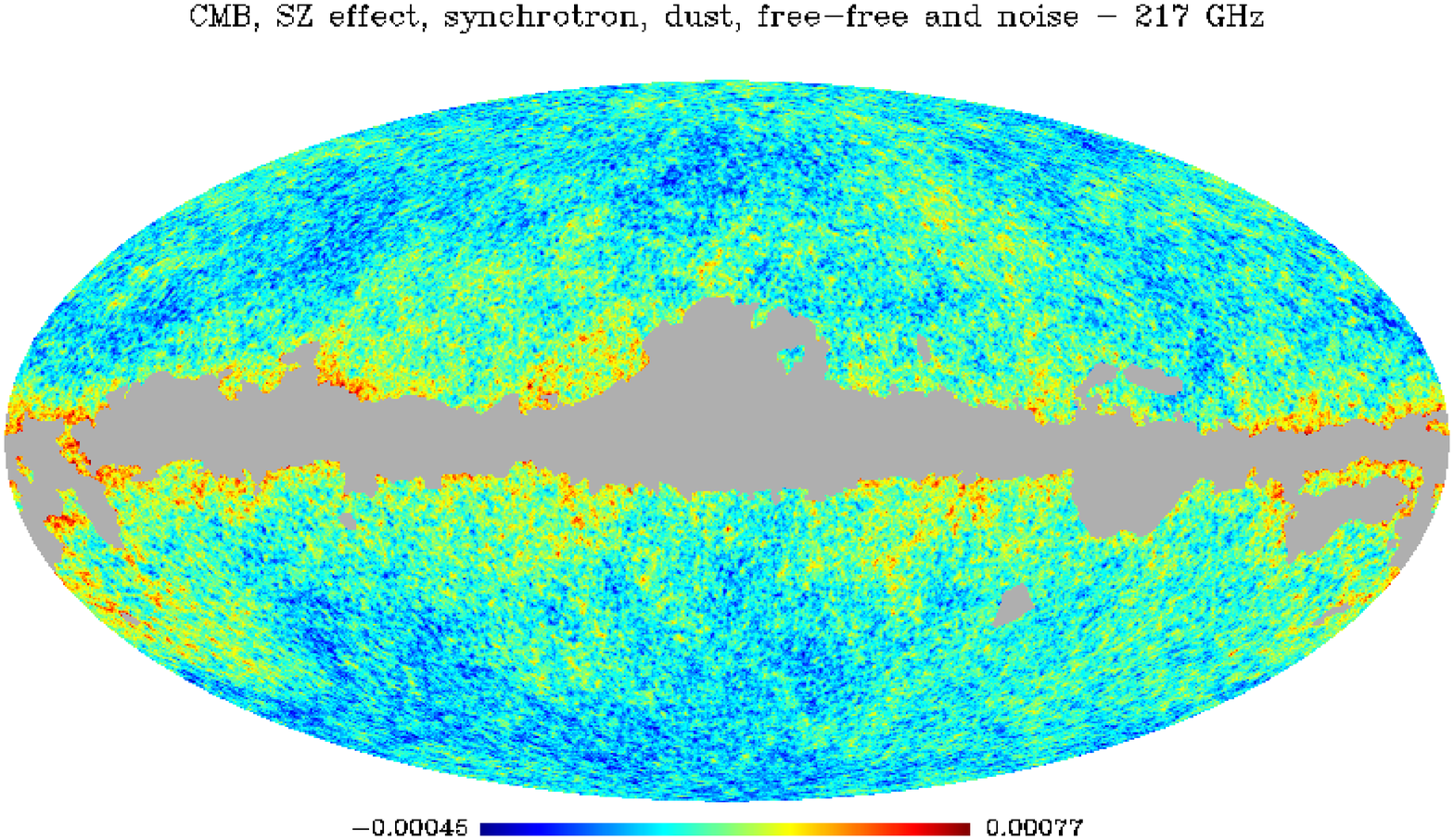}
\includegraphics[height=5cm]{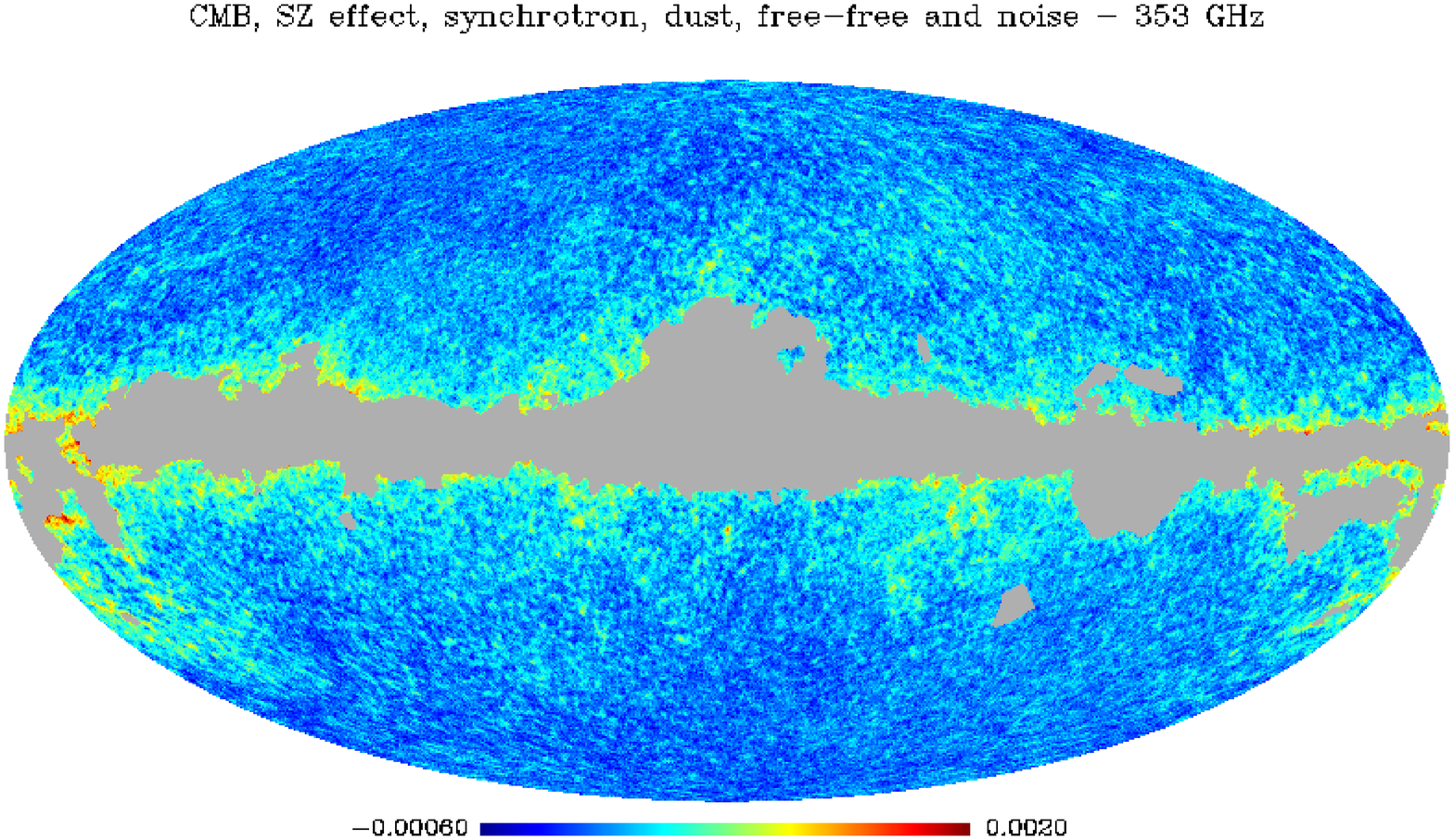}
\includegraphics[height=5cm]{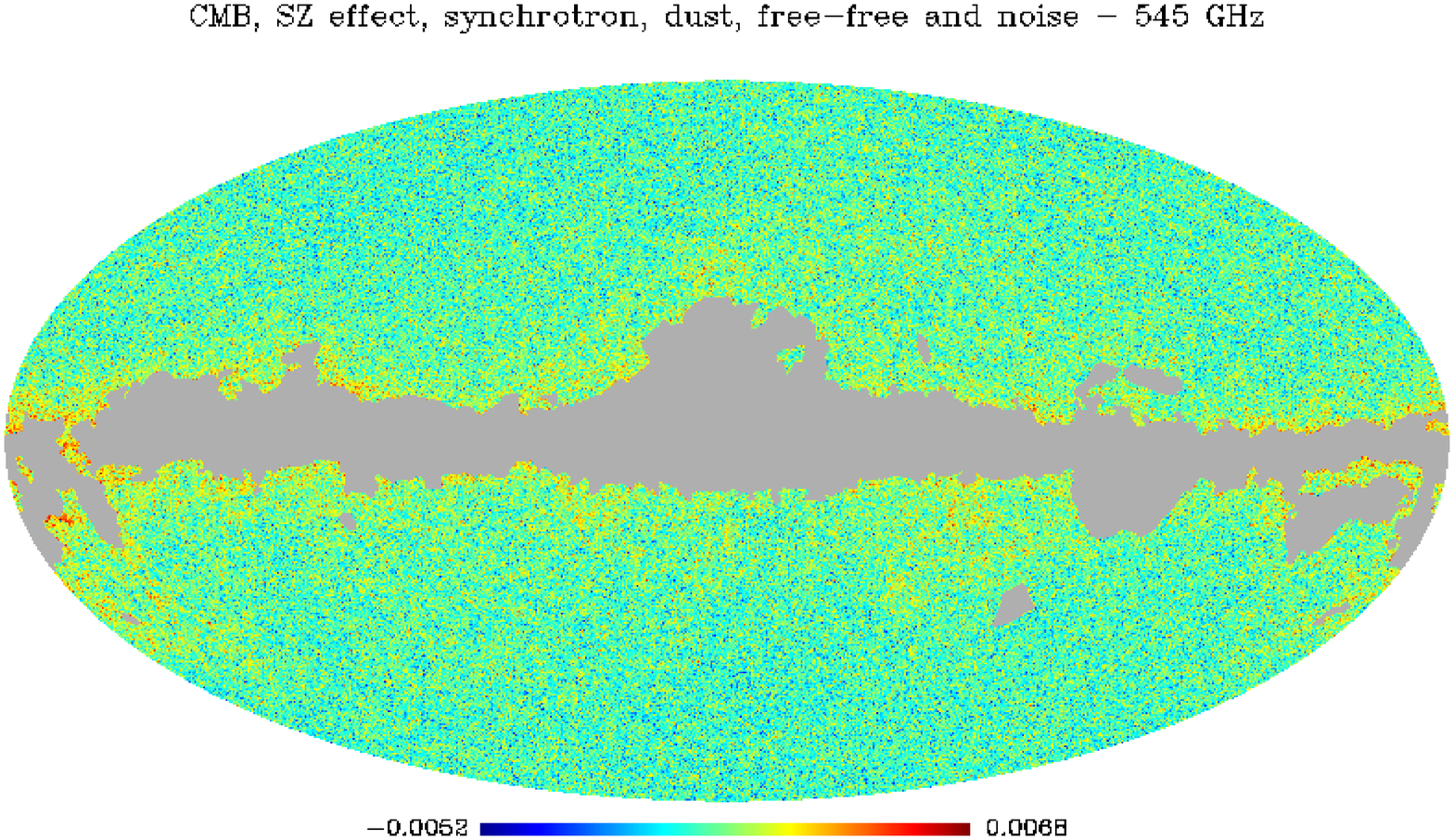}
\caption{Linear combination of CMB, SZ effect, Galactic emission (synchrotron, dust, and free-free) and HWGN maps. The unit of the maps is K, in Galactic coordinates and Mollweide projection.} \label{combinacao}
\end{figure*}

The SZ clusters were randomly placed all over the sky and we used the WMAP-7 KQ75 mask (also available at the LAMBDA website) to remove the Galactic plane neighbourhood, as usual in any CMB analysis. The resulting maps for the five \textit{Planck} HFI frequencies are shown in Fig. \ref{combinacao}.

\subsection{High-resolution full-sky simulations}\label{LAMBDA}

This second set of simulated maps was downloaded from the LAMBDA website. They have a $N_{side} = 8192$ pixelization, corresponding to a resolution of 0.4 arcminutes in six different frequencies: 148, 219, and 277 GHz (the ACT observing frequencies) and the additional 30, 90, and 350 GHz, close to the \textit{Planck} frequencies on the LFI and HFI. The maps are made of (1) the CMB affected by the lensing of an intervening structure between the last scattering surface and observers today; (2) the thermal and kinetic SZ effects, plus higher-order relativistic corrections, from galaxy clusters, groups, and the intergalactic medium; (3) a population of dusty star-forming galaxies that emit strongly at infrared (IR) wavelengths but still have significant microwave emission; (4) a population of galaxies, including active galactic nuclei, that emit strongly at radio wavelengths but still have significant microwave emission, and (5) the foreground emission of our own galaxy (dust, synchrotron, and free-free). A detailed explanation of these simulations can be found in \citet{2010/sehgal}.

The catalogue of SZ halos and both IR and radio galaxies included in these simulations are also available at the LAMBDA website.  The SZ catalogue contains 1414339 objects in the first octant, which are mirrored across the complete celestial sphere. The mass range is $2 \times 10^{11} M_{\odot} < M_{500} < 1.5 \times 10^{15} M_{\odot}$, with redshifts in the range $0 < z < 3$.

The simulated sky maps available at LAMBDA (hereafter LAMBDA maps) have a very fine resolution, and to avoid a very large computation time in analysing them, we re-pixelized them from $N_{side} = 8192$ to $N_{side} = 2048$. We convolved the lower-resolution maps with Gaussian beams with a FWHM extrapolated from the \textit{Planck} values (see Tab. \ref{bandas_lambda}), and then added noise.  Following the same procedure used in our ``homemade'' simulations,  two kinds of noise maps were used. One contained plain white Gaussian noise with a uniform coverage per pixel.  The second considered an asymmetric sky coverage, which was identical to the one described in Section \ref{simul_noise} (HWGN and NASC), but for which we used the white noise sensitivities given by the extrapolation of \textit{Planck} values in Tab. \ref{bandas}. These values are shown in Tab. \ref{bandas_lambda}.

\begin{table}[!h]
\centering 
\caption{Sensitivities extrapolated from \textit{Planck} frequencies.} {\footnotesize
\begin{tabular}{c c c c c c c}
\hline \hline
Frequency (GHz) & 30 & 90 & 148 & 219 & 277 & 350 \\
\emph{FWHM} (arcmin) & 32.65 & 9.42 & 6.73 & 4.66 & 4.43 & 4.44 \\
Sensitivity ($\mu K_{CMB} s^{1/2}$) & 146.8 & 25.7 & 14.2 & 20.9 & 32.5 & 74.3 \\
\hline
\end{tabular}} \label{bandas_lambda}
\end{table}


\section{Separation of components}\label{identification}

A CMB data set contains a combination of signals from many sources. The most significant come from our galaxy, the CMB itself, the SZ effect, and radio/IR point sources. 
Electronic noise is also produced by the detector and associated electronics. This section describes the method used to distinguish between the signals from these various components.

Our method is based on a numerical algorithm called the \emph{Joint Approximate Diagonalization of Eigenmatrices} (JADE) \citep{cardoso93, neural_comp} based on independent component analysis (ICA) \citep[see, e.g.,]{ica} and effective in extracting non-Gaussian components, as in the case of the SZ effect. We highlight its most interesting feature, that of not using any ``prior'' information about the input components. This feature sorts JADE from other methods used in the CMB/SZ analysis \citep{leach}.

The original JADE code is inefficient in the presence of noise and we introduced a wavelet pre-cleaning method prior to feeding the data to JADE.  After component separation, we used the SExtractor\footnote{http://www.astromatic.net/software/sextractor} package \citep{sextractor_artigo} to detect and identify the positions and intensities of the clusters. We describe below the steps of our pipeline, from the initial data preparation to the elaboration of the final catalogue of cluster candidates.  

The implementation of our pipeline was done fully in IDL (Interactive Data Language) for a number of reasons. First, this environment is very popular in the astronomy community, second, it is one of the languages used by the HEALpix package and, third, it is the chosen language of the CMB community for image processing. We modified the JADE routine available at MRS\footnote{http://irfu.cea.fr/en/Phocea/Vie\_des\_labos/Ast/ast\_visu.php?id\_ast =895} (Multi-Resolution on the Sphere) package, including a pre-whitening, wavelet-based, and processing step described in the next section.

\begin{figure*}[htb]
\centering
\includegraphics[width=16cm]{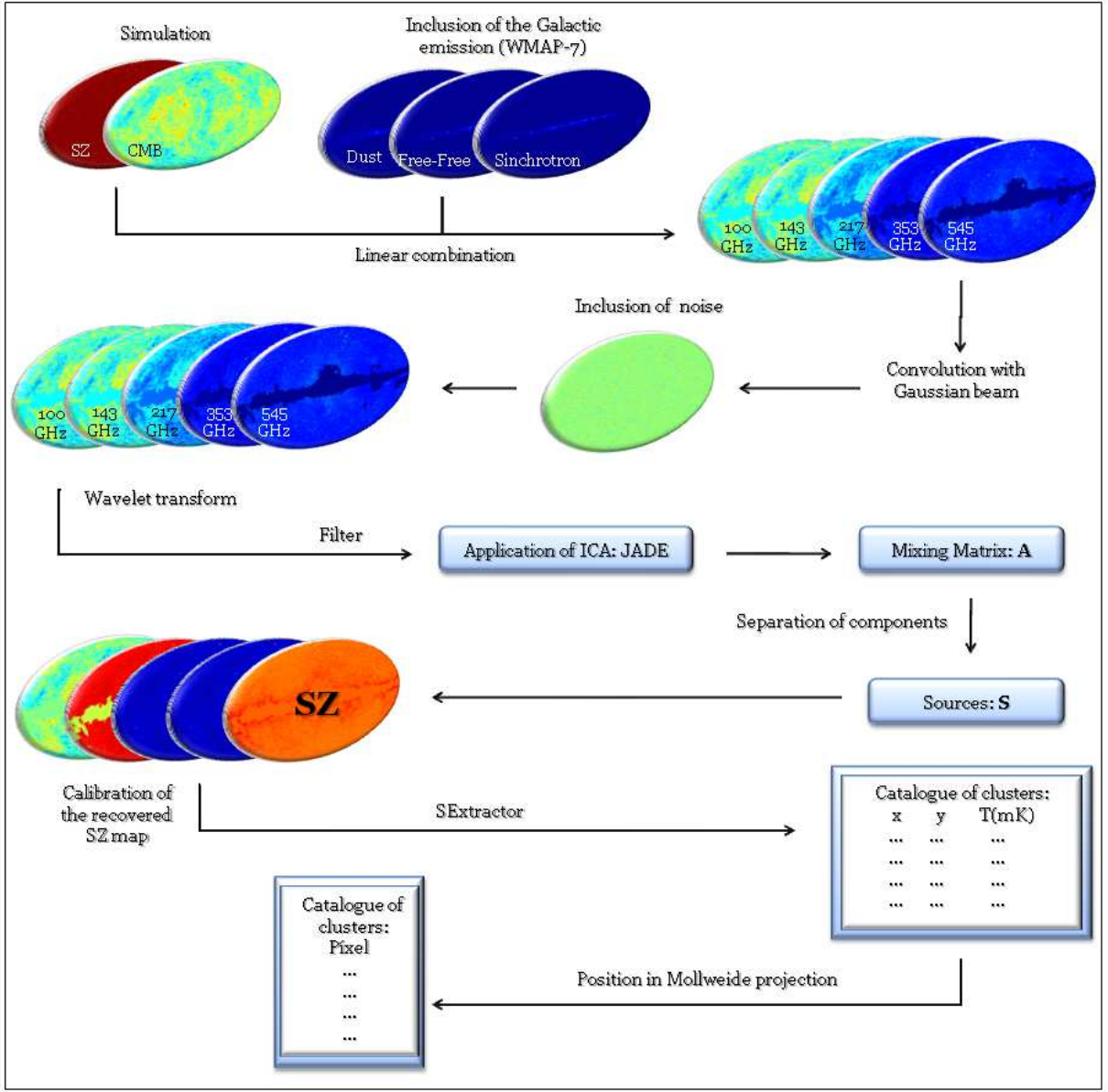}
\caption{Block diagram summarizing the SZ detection pipeline for simulated maps.} \label{diagram}
\end{figure*}

The processing time for the full cluster identification pipeline (pre-whitening+JADE+SExtractor) was $\sim$18 minutes for the ``homemade'' simulations and $\sim$ 1.2h for the LAMBDA maps.  Figure \ref{diagram} summarizes the data flow in our pipeline.

\subsection{Noise filtering}\label{denoising}

The presence of noise requires some sort of pre-processing to permit JADE to deal with the data. This pre-processing starts by wavelet-transforming each map. The transformation in wavelet space retains the information contained in the pixels while averaging the noise contribution and highlighting the data structures \citep{pires}.

We used the Daubechies wavelet transform to remove the noise from the data. The reasons for choosing this wavelet family are comprehensively discussed in, e.g., \citet{guide}. After conducting several tests, varying the order and level of the wavelet transform applied to the data, and comparing the results obtained with JADE in each test, we conclude that the best choices for this dataset is an order $N=3$ (db3) and a decomposition level $n=5$. It is important to remember that the higher the level used in the transformation, the more noise-free the data.

However, our various runs show there is an optimal decomposition level, above which a kind of ``saturation" occurs. When starting the denoising process, it is advisable to perform a few tests to verify the optimal level for a given dataset.

After the transformation of the data to the wavelet space, we filtered the maps with the HEALPix $smoothing.f90$ routine, using Gaussian beams with FWHM=8' for our ``homemade" simulations and FWHM=3' for the LAMBDA set to minimize the noise level prior to the application of JADE.

It is important to stress, at this point, that no previous information about the input data is used. This means that the cluster shape, mass thresholds, or redshift information, for instance, are not taken into account. Our wavelet tests are based solely upon the spectral information contained in the data.

\subsection{Sorting input signals: The JADE algorithm}

Many methods developed for signal separation are based on ICA, and can be considered a class known as \emph{blind source separation (BSS)} problems. A typical example of BSS is the processing of multidimensional data with no \emph{``a priori''} information \citep{ica/livro}.

This problem consists primarily of retrieving a set of $m$ statistically independent signals from $m$ mixtures of these instantly observed signals \citep[see, e.g.,][]{cardoso93, cardoso}. In other words, the goal is to estimate the matrix of the sources (independent components), $S$, and the mixing matrix, $A$, from $X$, the matrix of linear combinations of individual sources. This mixture model is described by the equation

\begin{equation}\label{BSS}
    X = A S,
\end{equation}

\noindent where $X$ is one $m \times T$ matrix, $T$ the number of observed samples (each row is a mixture of $m$ sources of a specific frequency), $S$ is a $m \times T$ matrix (each row is the signal from a particular source), and $A$ is a $m \times m$ invertible matrix, which specifies the original signal contributions of $S$ to $X$.

It is important to warn the user of some shortcomings and limitations of ICA \citep{ica/livro}. First, ICA assumes that the independent components are statistically independent. Second, at least one of the independent components must come from a non-Gaussian distribution, because Gaussian distributions have higher-order cumulants equal to zero, which mean that the ICA model cannot be applied. Finally, for the sake of simplicity, the model assumes that the mixed matrix is square, i.e., the number of independent components equals the number of observed mixtures. However, this is not a mandatory condition for using ICA; for details, we refer to \citet{ica/livro}.

In addition to these limitations, the ICA method does not return the actual amplitudes of signals, since these are initially unknown. However, this is not a major problem, since the signal can be recalibrated after the separation of the components. This issue is discussed in Section \ref{recovery}. In addition, the method does not allow the user to determine the sequential ordering of the independent components in the $S$ matrix rows, so the ordering can be freely changed.

Originally introduced by \citet{cardoso93}, JADE is a statistical, ICA-based, technique  that relies on high-order statistics. Its mixture model is given by Equation \ref{BSS} and assumes that the resulting sources in $S$ are non-Gaussian random processes with a high signal-to-noise ratio. Since a real noise-free map does not exist, there is a need for data pre-processing before applying this method.

We now describe the data processing steps used by JADE to obtain the independent components (i.e. the sources) \citep{ica/livro}. The method starts by centralizing data, assuming that both the mixture variables and the independent components have zero means, and it then performs a whitening of the observed signals. For the model described by Equation \ref{BSS}, the whitening of $X$ is carried out by the whitening matrix $V$ (the inverse of the square root of the covariance matrix of the data), generating the white vector $Z = VX = V A S$. We then compute a new orthogonal mixing matrix $W^T = V A$ and a new separation matrix $W$ \citep{ica/livro,pires}. The ICA Equation \ref{BSS} becomes $Z = W^T S$, after the whitening of the data.

The cumulant tensor of the whitened matrix $ Z $ has a special structure, which can be seen from the eigenvalue decomposition, that accounts for the independent components. To achieve this, the whole matrix is assumed to have the form $\mathbf{M}=\mathbf{w}_m \mathbf{w}_m^T$ (para $m~=~1,...,~n$) which is an eigenmatrix of the cumulant tensor

\begin{equation}
\mathbf{F}_{ij}(\mathbf{M}) = \lambda \mathbf{M}_{ij} = \sum_{kl} w_{mk} w_{ml} ~cum(z_i,z_j,z_k,z_l),
\end{equation}

\noindent where $w_m$ is a row of the $W$ matrix and $\lambda$ is the eigenvalue. Since the eigenvalues are distinct from each other, each eigenmatrix corresponds to an eigenvalue of the form $w_m w_m^T$, giving one of the rows of $W$. Thus, with knowledge of the eigenmatrices of the cumulant tensor it is possible to obtain the independent components. JADE was designed to solve the case for indistinguishable eigenvalues.

According to \citet{ica/livro}, the eigenvalue decomposition can also be understood as a diagonalization process. Hence, the eigenvalue decomposition is also a diagonalization of the cumulant tensor $\mathbf{F}(\mathbf{M})$ that is performed by multiplying the matrix $W$ for any $\mathbf{M}$, as

\begin{equation}
Q = \mathbf{W}\mathbf{F}(\mathbf{M_i})\mathbf{W}^T.
\end{equation}

\noindent Thus, $M_i$ is chosen such that $Q$ is as diagonal as possible.

Since the $W$ matrix is orthogonal, its multiplication by another matrix does not change the total sum of squares of the elements of this matrix, thus minimizing the sum of the squares of the off-diagonal elements is equivalent to maximizing the sum of the squares of the diagonal elements. Thus, this algorithm aims to maximize the equation

\begin{equation} \label{jade/diag}
\mathfrak{I}_{JADE}(\mathbf{W}) =\sum \limits_i \| diag (\mathbf{W}\mathbf{F}(\mathbf{M_i})\mathbf{W}^T) \|^2.
\end{equation}

The maximization of $\mathfrak{I}_{JADE}$ is a method of the joint approximate diagonalization of $\mathbf{F}(\mathbf{M}_i)$. The $M_i$ matrices are chosen from the eigenmatrices of the cumulant tensor, which provide all relevant information about the cumulants because they share the same space as the cumulant tensor.

The $A$ matrix is obtained by applying JADE to the data in wavelet space. Multiplying its inverse ($A^{-1}$) by $X$, we obtain the $S$ matrix of components. This result can be achieved because the application of wavelet transform does not affect the $A$ matrix, but only increases the accuracy of the calculation. Since the $A$ matrix was carefully calculated, it was applied to the input data to extract the SZ map.

Figure \ref{sz_rec} shows an example of the extraction of an SZ map, obtained from the analysis of the ``homemade" input maps with HWGN. It can be seen from these results that the temperature scale of the recovered SZ map does not match the scale of the simulated maps, since JADE loses the calibration information during data processing. The next section discusses the process of calibration recovery for each frequency.

\subsection{Recovering calibration}\label{recovery}

The appropriate method for calibrating the recovered map is derived from an initial analysis comparing the output map to the input map to see how the fluxes of known clusters or other potential calibrators change in each map region. We found that the intensity of the recovered map by JADE differs from the input data by a nearly constant value across the whole map. Since we do not deal with real data, no known sources can be used to reconstruct the calibration, so we used our fake input clusters to accomplish this task. In a real map, however, prior knowledge of the fluxes of a few well-known sources allows the calibration for the full map to be made.

We took the clusters' central values $\Delta T_{SZ}$ in the input and output maps and calculated the ratio of both for each selected cluster. The average value of these ratios, at each of the frequencies, is the value by which the recovered map is multiplied to recover the calibration. In this work, we used 50 clusters randomly chosen to perform the procedure, since the larger the number of clusters used in the calibration, the more accurate the result.

\begin{figure*}[t!]
\begin{minipage}[t]{1.\linewidth}
\centering
\includegraphics[width=14cm]{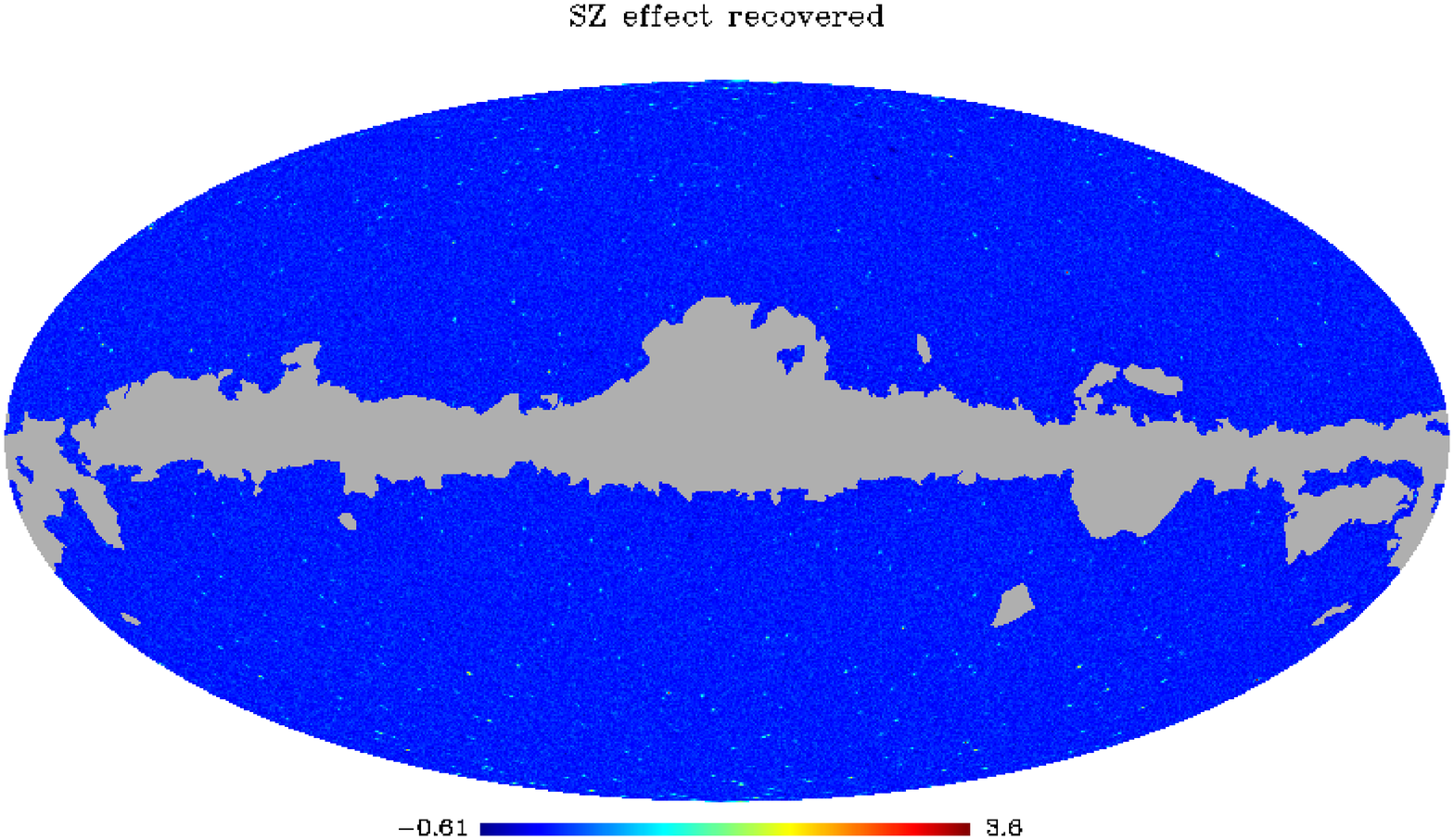}
\caption{Our SZ-effect map recovered with JADE algorithm from the analysis of our ``homemade" maps contaminated by HWGN.} \label{sz_rec}
\end{minipage}

\begin{minipage}[t!]{1.\linewidth}
\centering
\includegraphics[height=7cm]{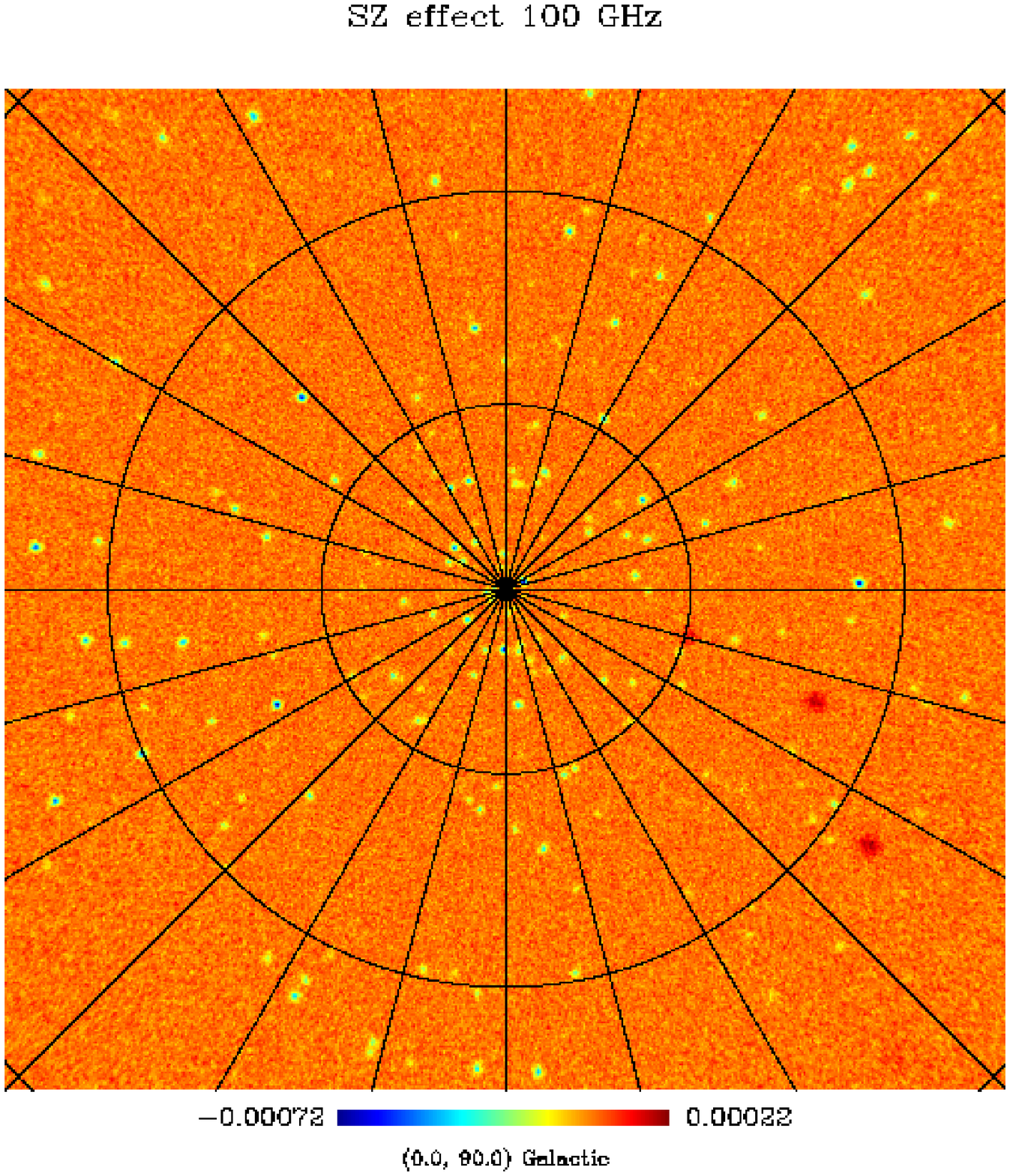}
\includegraphics[height=7cm]{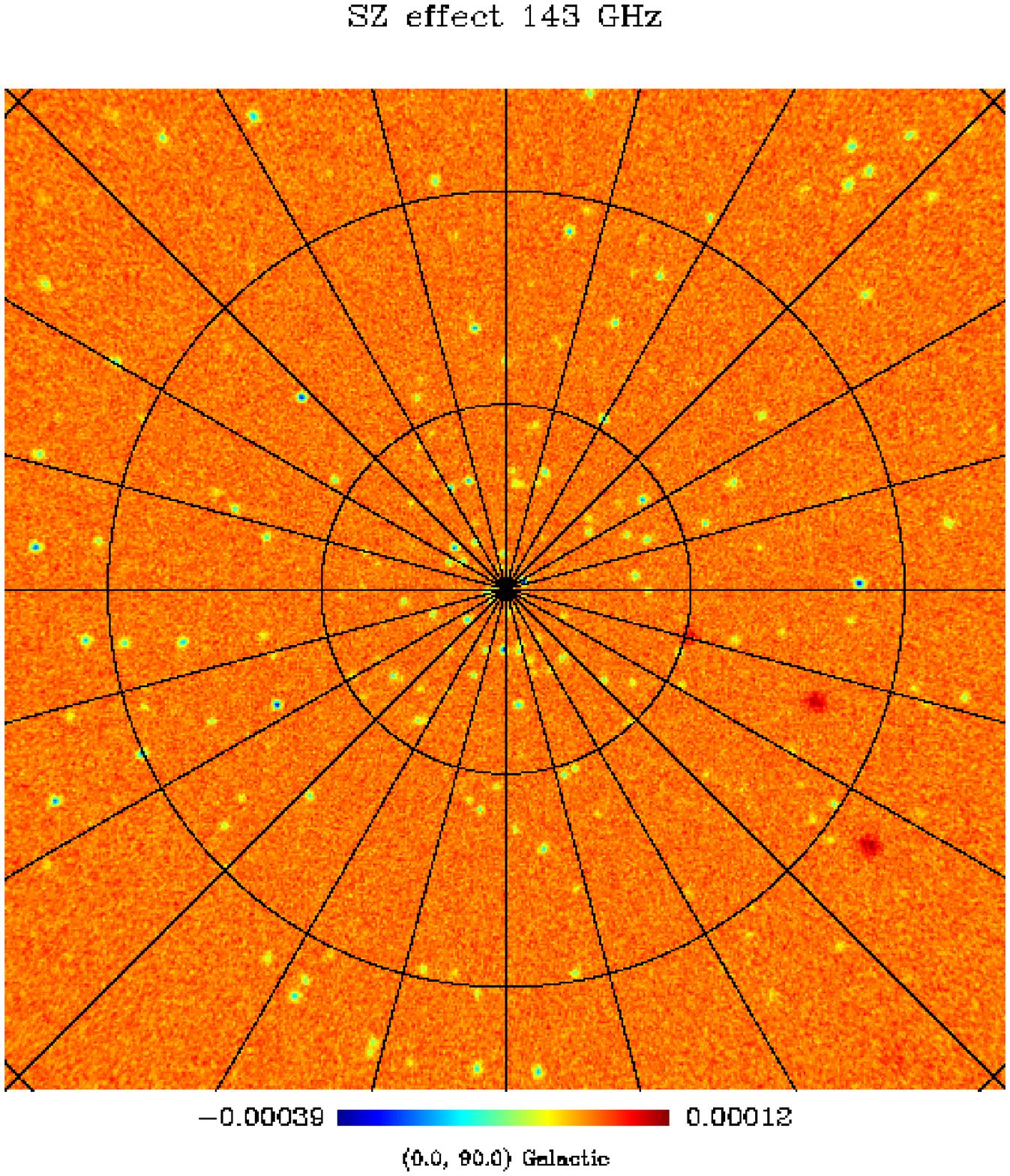}
\includegraphics[height=7cm]{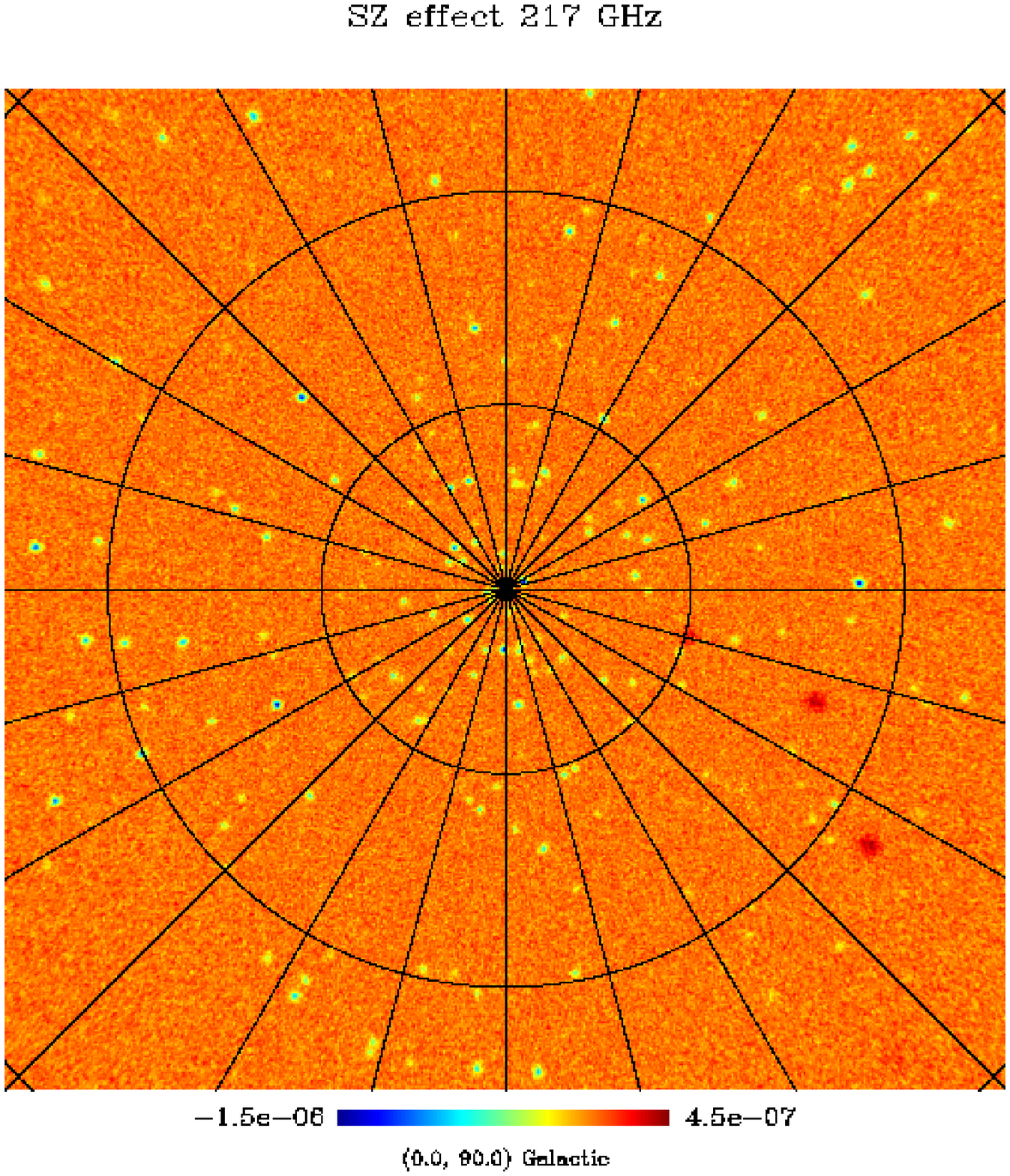}
\includegraphics[height=7cm]{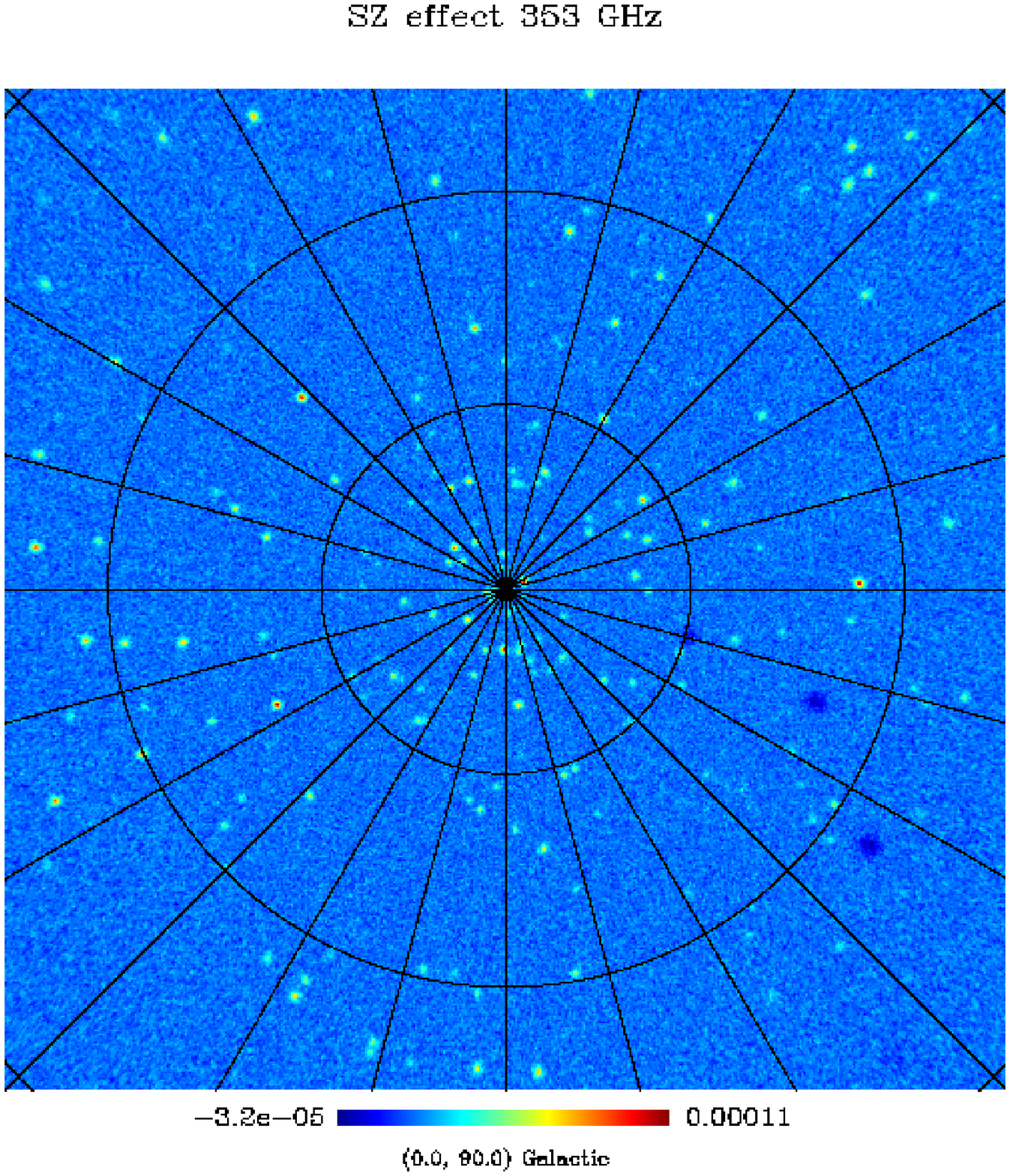}
\includegraphics[height=7cm]{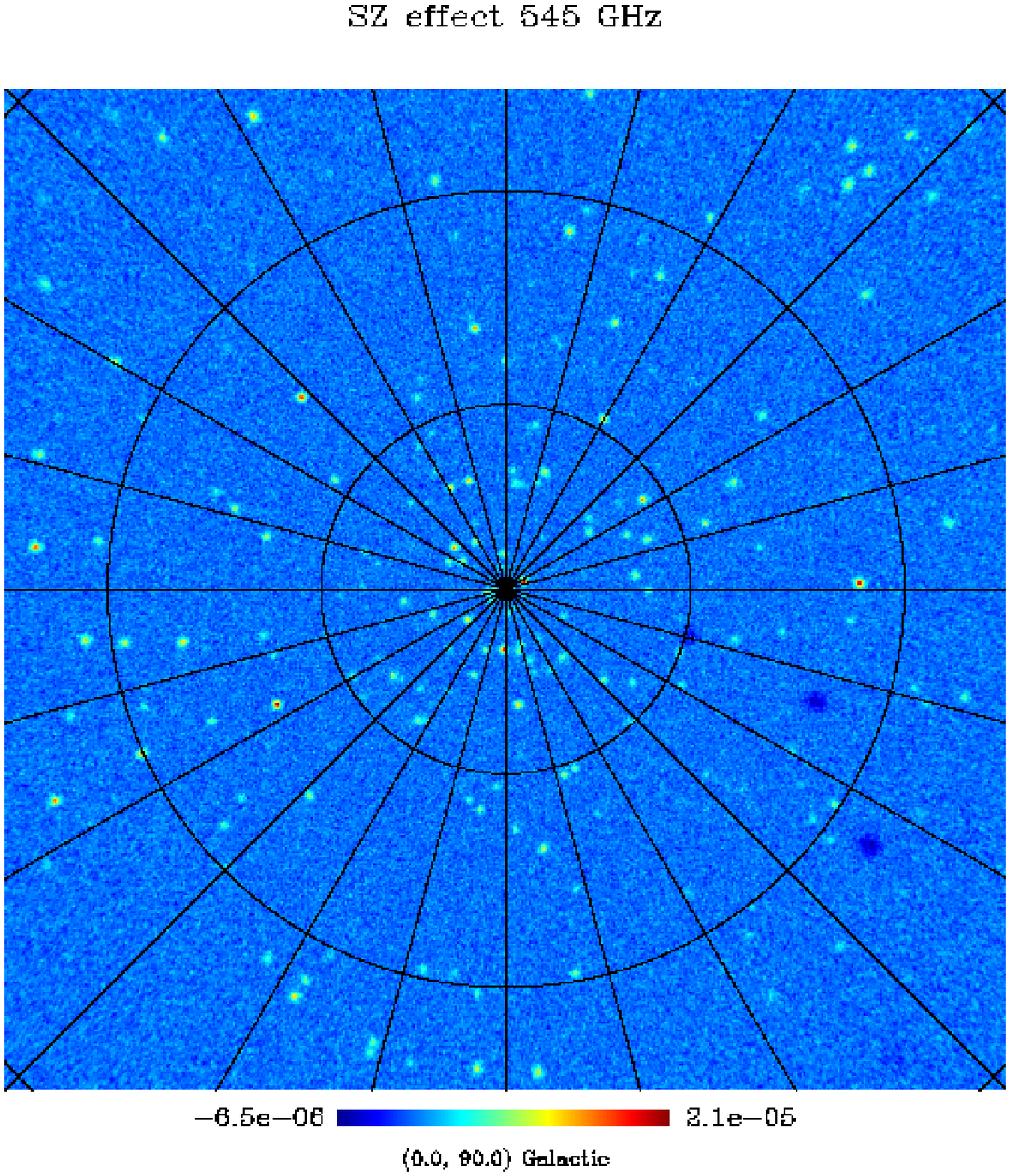}
\caption{Gnomonic projection centred on the north Galactic pole, of the SZ map in Fig. \ref{sz_rec} (``homemade" + HWGN result) when calibrated for each input frequency.} \label{sz_calib}
\end{minipage}
\end{figure*}

A section of the map in Fig. \ref{sz_rec} calibrated for each frequency is shown in Fig. \ref{sz_calib}, which can be inspected and visually compared to the same section of the simulated map (Fig. \ref{sz_simul}) to check for large differences in the temperature scales.

\begin{figure}[th!]
\resizebox{\hsize}{!}{\includegraphics{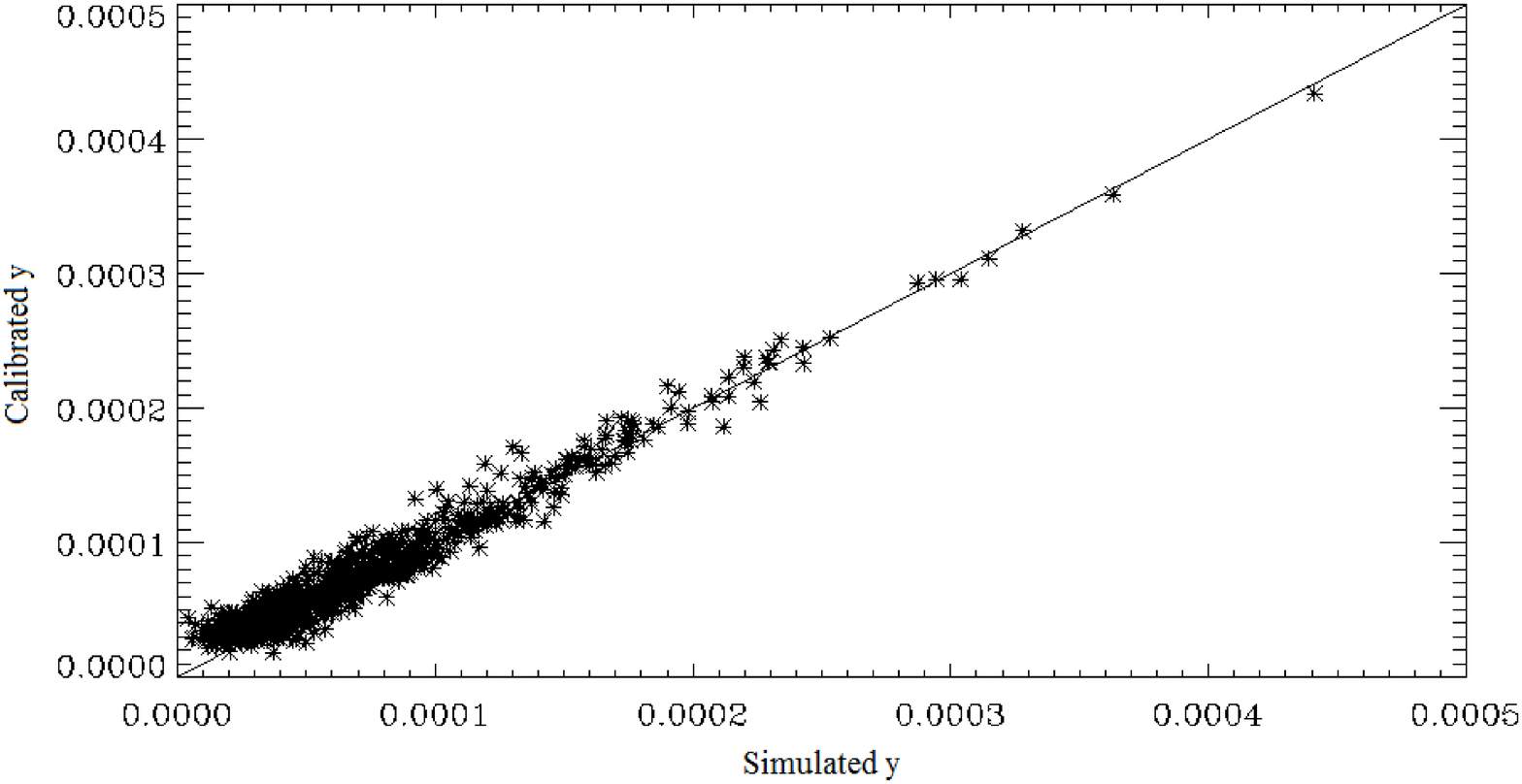}}
\caption{Graph of simulated and calibrated $y$ parameters, obtained analysing our ``homemade" simulation with HWGN. The diagonal straight is the line of equality.} \label{graf_calib}
\end{figure}

We proceeded to compare the $y$ values of the profile amplitude (the central value) calculated from both the simulated and calibrated $\Delta T_{SZ}$ values of the clusters, shown in Fig. \ref{graf_calib}. In this graph, each point is equivalent to a single cluster and the diagonal lines represent the ``equality line''. The closer the point to the diagonal, the closer the input (simulated) and output (recovered and calibrated) $y$ values.  Thus, the plot in Fig. \ref{graf_calib} is a good indicator of accuracy of our calibration method for the JADE output maps. We also estimate the average dispersion $D$ of the data

\begin{equation} \label{dispersion}
\textrm{D} = {1 \over N_{pts}}\sum_{i=1}^{N_{pts}} \Bigg | {y^{sim}_i - y^{cal}_i \over y^{sim}_i} \Bigg |,
\end{equation}

\noindent obtaining 0.27.


\section{Cluster detection}\label{results}

After recovering the clusters from a ``full sky with noise" using JADE, and recalibrating their fluxes, we used the SExtractor package \citep{sextractor_artigo,sextractor_dummies} to select the cluster candidates.  The most important SExtractor parameters, and those that most strongly influence the results, are DETECT\_THRESH (the detection threshold), DETECT\_MINAREA (which sets the minimum number of pixels above the threshold triggering detection), and FILTER\_NAME (which selects the file containing the filter definition).

SExtractor offers a number of filters to be used in this kind of analysis, that have various FWHM and sizes (both given in pixels). For our analysis, the  filter that most closely recovered the input data was the Gaussian filter. For the ''homemade'' datasets, we used a FWHM of $4$ pixels and a mask with $7 \times 7$ pixels. For the LAMBDA dataset, the values were, respectively, 2 and $5 \times 5$ pixels. In addition, we used threshold values of $2.5\sigma$, $1.5\sigma$, $2.0\sigma$, and $2.0\sigma$ for the ``homemade" datasets + HWGN, the ``homemade" datasets + NASC, the LAMBDA datasets + HWGN, and the LAMBDA datasets + NASC, respectively. DETECT\_MINAREA was set equal to $4$ and $8$ for the ``homemade" and LAMBDA datasets.

We compare the positions of the calibrated clusters found by SExtractor with those included in the input sky maps, to account for false detections.
The criterion used to make that determination was that for each position of cluster candidate indicated by SExtractor we checked for the existence of clusters in a circle, of radius equal to three pixels, around that position. If there was no cluster in the region, the candidate was considered a false detection. If there was more than one, the most massive cluster was assumed to be the detection,  since it is more likely that one finds the most massive cluster. At this point, we did not consider the possibility of multiple detections.

\begin{figure}[th!]
\centering
\resizebox{\hsize}{!}{\includegraphics{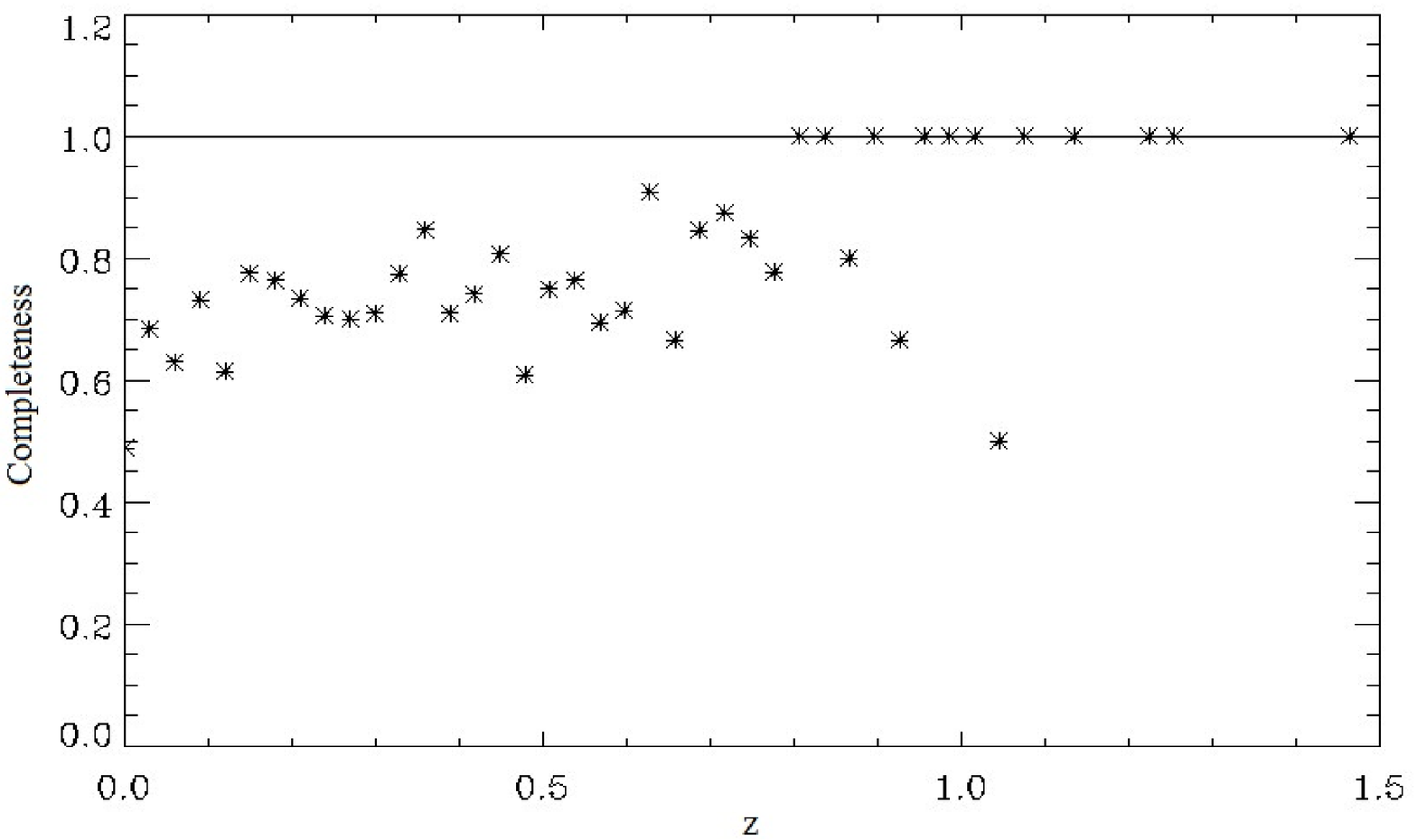}}
\resizebox{\hsize}{!}{\includegraphics{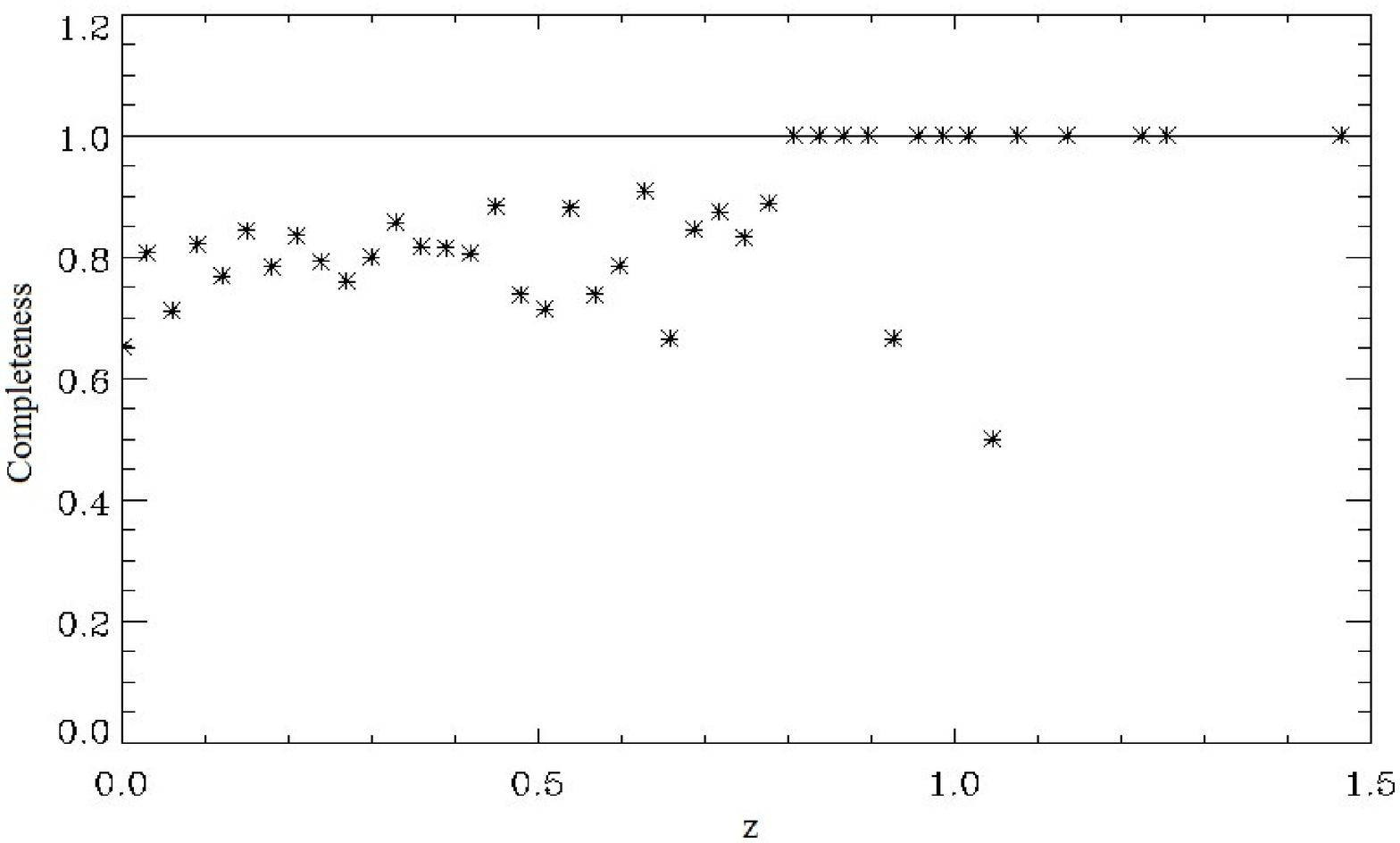}}
\resizebox{\hsize}{!}{\includegraphics{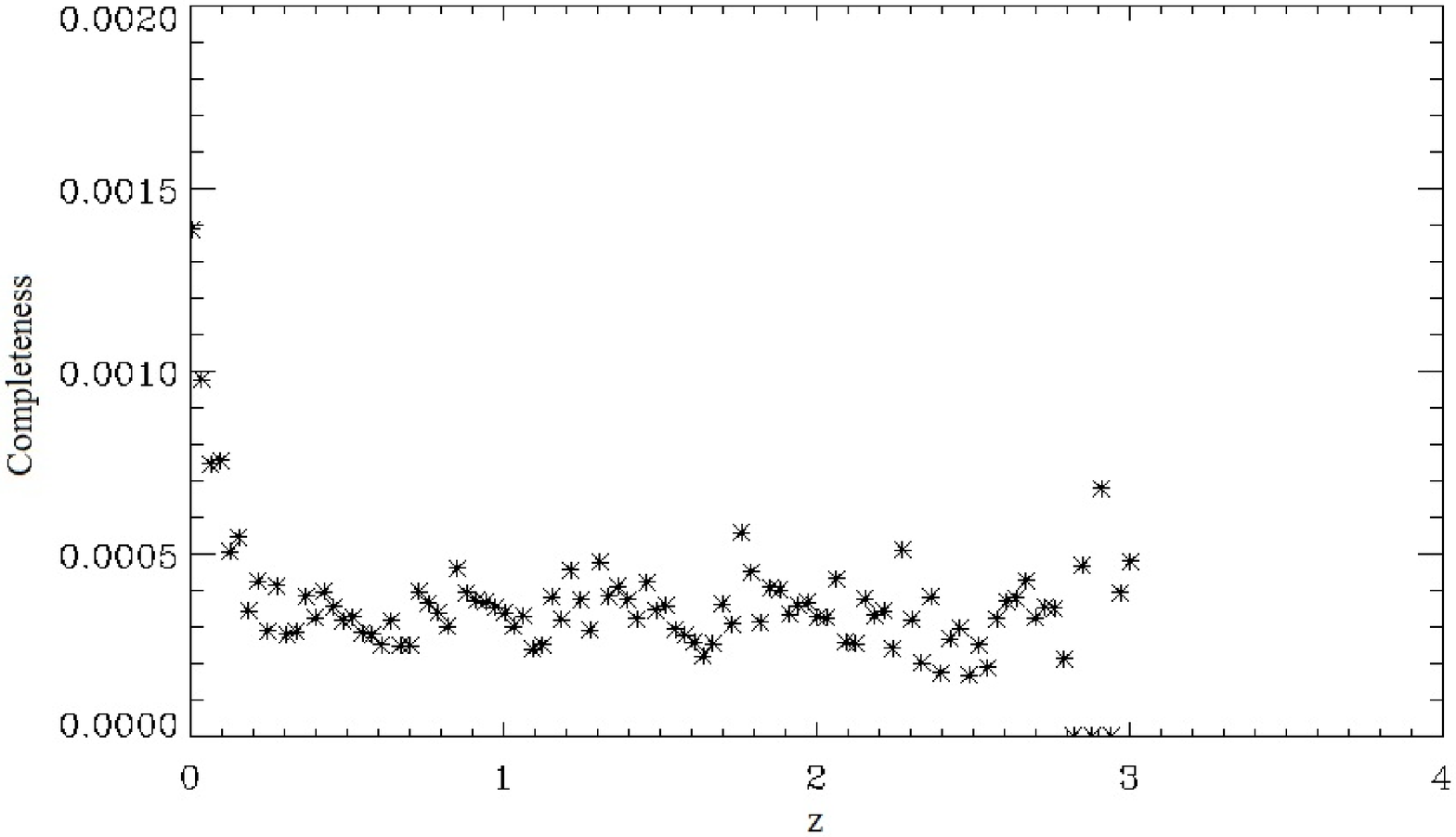}}
\resizebox{\hsize}{!}{\includegraphics{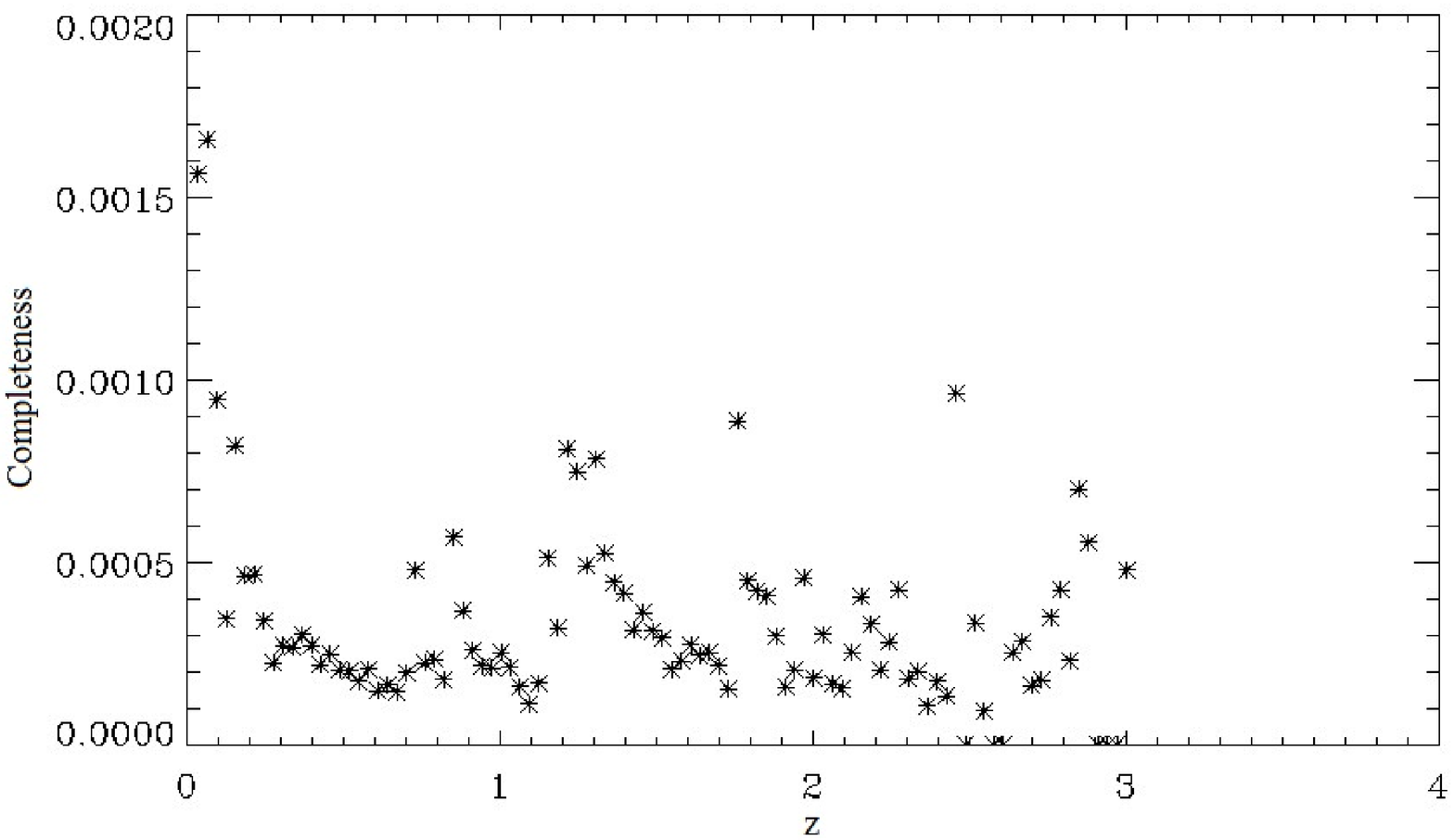}}
\caption{Relation between completeness and redshift intervals for recovered SZ catalogues. The graphics from top to down correspond to results of the analysis of ``homemade" + HWGN, ``homemade" + NASC, LAMBDA + HWGN and LAMBDA + NASC, respectively.} \label{completeness_z}
\end{figure}

\begin{figure}[th!]
\centering
\resizebox{\hsize}{!}{\includegraphics{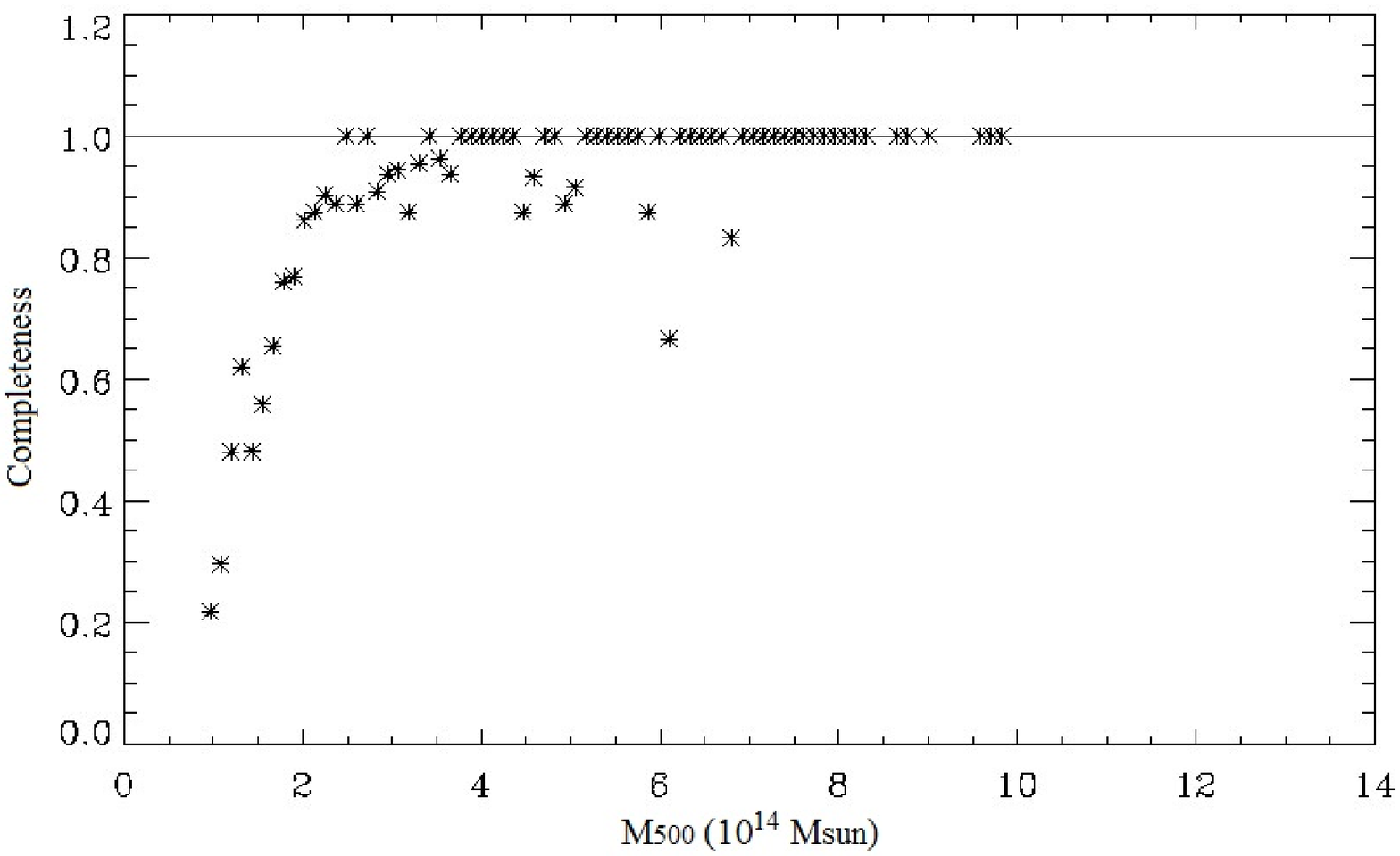}}
\resizebox{\hsize}{!}{\includegraphics{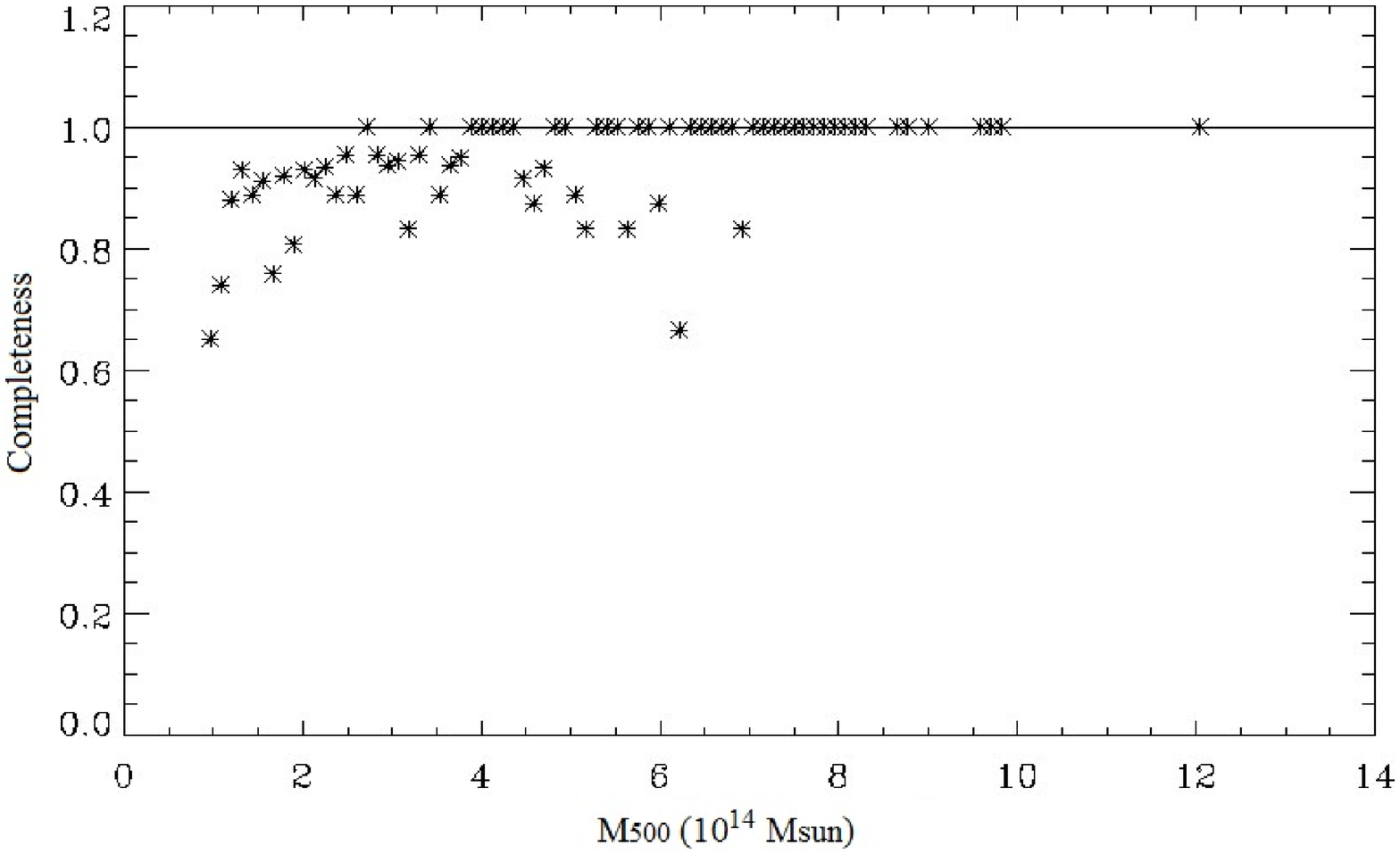}}
\resizebox{\hsize}{!}{\includegraphics{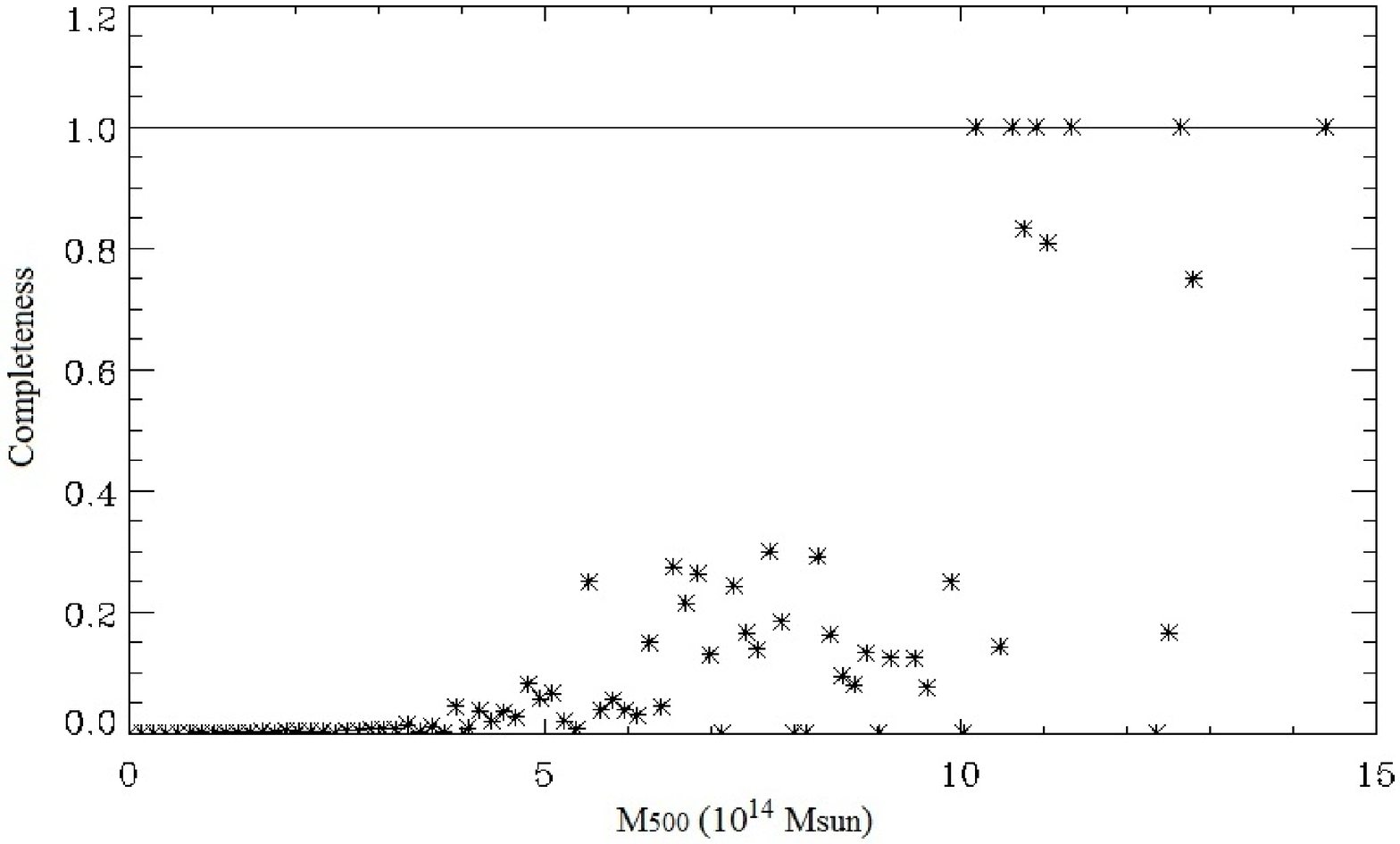}}
\resizebox{\hsize}{!}{\includegraphics{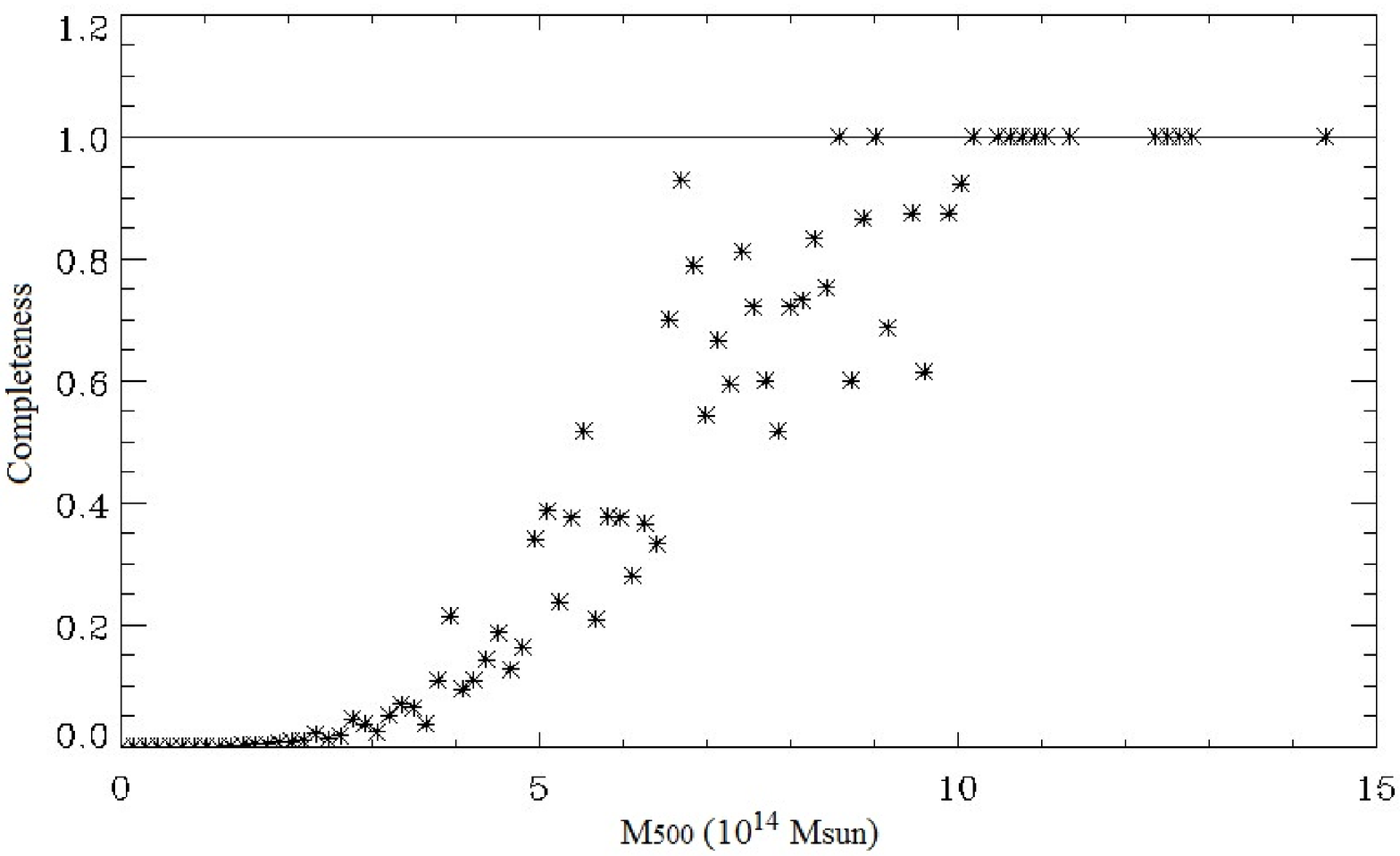}}
\caption{Relation between the completeness and $M_{500}$ intervals for the recovered SZ catalogues. The graphics from top to down correspond to the results of the analysis of ``homemade" + HWGN, ``homemade" + NASC, and LAMBDA + HWGN and LAMBDA + NASC datasets, respectively.} \label{completeness_m500}
\end{figure}

Our results obtained from the analysis of both datasets and noise types are summarized in Tab. \ref{results_cases}, which shows the number of cluster candidates indicated by SExtractor, the number of confirmed clusters, and finally both the purity and completeness of the recovered ``catalogue". The Figs. \ref{completeness_z} and \ref{completeness_m500} present the completeness by redshift and mass interval for each dataset. The first one shows that the completeness does not change significantly with redshift, which highlights the redshift independence of the SZ effect, as expected. The second one shows the sensitivity of the SZ effect to the mass of the cluster, the completeness increasing with increasing mass. It can also be seen from these figures that the different noise models led to slightly different results, implying that one has to test the pipeline parameters to find the most appropriate filtering scheme (with respect to instrumental properties such as beam size and expected noise level) for a given dataset.

\begin{table*}[th!]
\begin{minipage}[t]{1.\linewidth} 
\begin{center}
\caption{Results for both datasets.}
\label{results_cases}
\begin{tabular}{ccccc}
 \hline\hline
Input maps & Found clusters & Confirmed clusters & Purity\tablefootmark{a} (\%) & Completeness\tablefootmark{b} (\%)\\
 \hline
``Homemade" + HWGN & 766  & 725  & 94.6 & 72.5 \\
``Homemade" + NASC      & 813  & 804  & 98.9 & 80.4 \\
LAMBDA + HWGN      & 3928 & 3873 & 98.6 & 0.034 \\
LAMBDA + NASC           & 3560 & 3550 & 99.7 & 0.031 \\
  \hline
\end{tabular}
\end{center}
\tablefoottext{a}{$purity = true ~ detections / total ~ detection$}\\
\tablefoottext{b}{$completeness = true ~ detections / simulated ~ clusters$}
\end{minipage}
\end{table*}

There is a considerable difference between the two datasets used in this work, which allowed us to test and evaluate the pipeline under two very different conditions.

The total completeness obtained from the analysis of our simple ``homemade" simulations provide a first indication of the efficiency of the method. However, it is insufficient, since it does not account for other contaminants of the SZ signal, such as the cross-terms of the thermal and kinetic SZ effects, radio sources, unresolved SZ clusters, and the SZ background. A more thorough testing was done by processing the LAMBDA maps through the pipeline. Despite the large differences between both datasets, we obtained a very similar result.

Another point worth mentioning is that, in this work, it makes no sense to discuss the total completeness of the resulting catalogue, since there is a very large number of clusters distributed over the full sky, most of them well below $5 \times 10^{13} M_{\odot}$, and so unresolved via the SZ signal. Nevertheless, the results obtained using the LAMBDA maps presented a level of purity above $90 \%$. The completeness of the mass and redshift intervals behave as expected, reaching a very low levels at lower masses and clearly increasing towards tens of percent for masses above $5 \times 10^{14} M_{\odot}$.


\section{Concluding remarks}\label{closing}

We have presented our implementation of a method to identify galaxy clusters using the SZ effect in CMB observations.  We have adapted JADE, a publicly available algorithm, to deal with noisy data by applying a pre-whitening, wavelet-based process and added a source detection package (SExtractor) at the end of our pipeline.

We have found the most attractive feature of this method is that it is based on a blind search algorithm, i.e., its application \emph{really} does not require any \emph{a priori} information about the targets. The essential contributions of this work were the following:

1) The wavelet-based analysis tool has been adapted to perform the initial cleaning of the input data. Since JADE was designed to perform in the absence of noise, data preprocessing was essential to ensure the efficient performance of our algorithm.

2) A parameter set was determined for the full pipeline (wavelet tool, JADE, and SExtractor) that delivered catalogues from two simulated datasets with a level of purity (ratio of confirmed clusters to total detected clusters) above 90\%.

Our method is a complementary approach to the MF algorithm currently used by the \textit{Planck}, SPT and ACT collaborations \citep{2011/planck_results_cluster,2011spt_story,2011act_hand}, and can be used as a redundant tool in their data analysis pipeline.
The results of using MF in CMB data analysis are widely described in the literature, hence we perform no further testing here. Our goal is to describe an alternative (and useful) technique to identify SZ clusters with \textit {no prior assumptions about the input data} and under very different input conditions. We do not intend to perform a direct comparison between the two methods.

We have developed a full pipeline\footnote{The full package, with routines and instructions, will be available at http://www.das.inpe.br/$\sim$alex/SZ/SZHunter\_pipeline.tar.gz.}, which is represented in the block diagram of Fig. \ref{diagram} and can be summarized in four main steps: data preprocessing (de-noising) based on a wavelet tool, separation of components (emissions) by JADE, calibration of the recovered SZ map, and the identification of the positions and intensities of the clusters using the SExtractor package.
Two simulated datasets were run through this pipeline: a ``homemade'' set and the more complete LAMBDA dataset, which were both described in Sections \ref{homemade} and \ref{LAMBDA}.

The results presented in Tab. \ref{results_cases} of Section \ref{results} indicate that our method performed very efficiently for both datasets. They vary slightly according to the characteristics of the data, especially in terms of the noise characteristics, and we caution that the whole pipeline may perform differently when applied to real data. 
Thus, the application of our method to real data may require some adjustment in the preprocessing phase to determine the optimal parameters for the denoising and target extraction, as discussed in Sections \ref{denoising} and \ref{results}.

\begin{acknowledgements}
We are grateful to J. F. Cardoso for releasing the JADE code on web, to E. Komatsu for the routine adapted and used in the SZ effect profile simulations (available at http://gyudon.as.utexas.edu/$\sim$komatsu/CRL/index.html), and Moudden et al. for the MRS package. We thank the referee for the suggestions and, particularly, the insightful comments about our noise estimation. We also thank Rodrigo Leonardi for the discussions about the implementation of the pipeline and many interesting suggestions for testing it. We acknowledge the HEALPix collaboration for providing the pixelisation scheme used in this work and the LAMBDA website for storing and making available the WMAP data and the online version of CMBFAST. C.A. Wuensche acknowledges the CNPq grant 2010/300400-3.
\end{acknowledgements}



\end{document}